\documentclass[reqno]{amsart}

\usepackage{graphicx}
\usepackage{caption}
\usepackage{subcaption}
\usepackage{booktabs}
\usepackage[table,xcdraw]{xcolor}
\usepackage{rotating,tabularx,siunitx}
\newcolumntype{T}{>{\raggedleft\arraybackslash} X}

\usepackage[implicit=false]{hyperref}

\usepackage{pdflscape}
\usepackage{tikz}
\usepackage[ruled,linesnumbered]{algorithm2e}
\usepackage{IEEEtrantools}
\usepackage{amssymb,latexsym,amsfonts,amsmath,mathrsfs}
\usepackage{dsfont}

\usepackage{amsthm}
\usepackage{amsmath}
\DeclareMathOperator*{\argmax}{argmax}
\DeclareMathOperator*{\argmin}{argmin}
\def\reals{\mathbb{R}}
\def\nats{\mathbb {N}}

\newcommand{\until}{\mathbin{\sf U}}
\def\borel{\mathscr{B}}

\usepackage{listings}
\usepackage{xcolor}
\usepackage{enumerate}
\definecolor{codegreen}{rgb}{0,0.6,0}
\definecolor{codegray}{rgb}{0.5,0.5,0.5}
\definecolor{codepurple}{rgb}{0.58,0,0.82}
\definecolor{backcolour}{rgb}{0.95,0.95,0.92}
\lstdefinestyle{mystyle}{
	backgroundcolor=\color{backcolour},   
	commentstyle=\color{codegreen},
	keywordstyle=\color{magenta},
	numberstyle=\tiny\color{codegray},
	stringstyle=\color{codepurple},
	basicstyle=\ttfamily\footnotesize,
	breakatwhitespace=false,         
	breaklines=true,                 
	captionpos=b,                    
	keepspaces=true,                 
	numbers=left,                    
	numbersep=5pt,                  
	showspaces=false,                
	showstringspaces=false,
	showtabs=false,                  
	tabsize=2
}

\usepackage{arydshln}

\SetKwInput{KwData}{Input}
\SetKwInput{KwResult}{Output}

\newtheorem{theorem}{Theorem}[section]

\newtheorem{definition}[theorem]{Definition}

\newtheorem{remark}[theorem]{Remark}

\numberwithin{equation}{section}

\usepackage{fancyhdr}

\newenvironment{nouppercase}{%
	\renewcommand{\uppercasenonmath}[1]{}}{}

\begin{document}

\begin{abstract}
This paper is concerned with developing a software tool, called \textsf{IMPaCT}, for the parallelized verification and controller synthesis of large-scale stochastic systems using interval Markov chains (IMCs) and interval Markov decision processes (IMDPs), respectively. The tool serves to (i) construct IMCs/IMDPs as finite abstractions of underlying original systems, and (ii) leverage \emph{interval iteration} algorithms for formal verification and controller synthesis over \emph{infinite-horizon} properties, including safety, reachability, and reach-avoid, while offering \emph{convergence guarantees}. \textsf{IMPaCT} is developed in C++ and designed using AdaptiveCpp, an independent open-source implementation of SYCL, for adaptive parallelism over CPUs and GPUs of all hardware vendors,
including Intel and NVIDIA. \textsf{IMPaCT} stands as the first software tool for the parallel construction of IMCs/IMDPs, empowered with the capability to leverage high-performance computing platforms and cloud computing services. Specifically, parallelism offered by \textsf{IMPaCT} effectively addresses the challenges arising from the state-explosion problem inherent in discretization-based techniques applied to large-scale stochastic systems. We benchmark \textsf{IMPaCT} on several physical case studies, adopted from the ARCH tool competition for stochastic models, including a $2$-dimensional robot, a $3$-dimensional autonomous vehicle, a $5$-dimensional room temperature system, and a $7$-dimensional building automation system. To show the scalability of our tool, we also employ \textsf{IMPaCT} for the formal analysis of a $14$-dimensional case study.\\

\end{abstract}

\title{{\LARGE \textbf{\textsf{IMPaCT}}: \underline{\textbf{I}}nterval \underline{\textbf{M}}DP \underline{\textbf{Pa}}rallel Construction for \underline{\textbf{C}}ontroller Synthesis of Large-Scale S\underline{\textbf{T}}ochastic Systems}}

\author{{\bf {\large Ben Wooding \and Abolfazl Lavaei}}\\{\normalfont School of Computing, Newcastle University, United Kingdom}\\ \textsf{\{ben.wooding,abolfazl.lavaei\}@newcastle.ac.uk}}

\pagestyle{fancy}
\lhead{}
\rhead{}
  \fancyhead[OL]{Ben Wooding and Abolfazl Lavaei}

  \fancyhead[EL]{\textbf{\textsf{IMPaCT}}: IMDP Parallel Construction for Controller Synthesis of Large-Scale Stochastic Systems}
  \rhead{\thepage}
 \cfoot{}
 
\begin{nouppercase}
	\maketitle
\end{nouppercase}

\section{Introduction}
Large-scale stochastic systems serve as a crucial modeling framework for characterizing a wide array of real-world safety-critical systems, encompassing domains such as power grids, autonomous vehicles, communication networks, smart buildings, energy systems, and so on. The intended behavior of such complex systems can be formally expressed using high-level logic specifications, typically formulated as \emph{linear temporal logic (LTL)} expressions~\cite{katoen2008book}. Automating the formal verification and controller synthesis for these complex systems that fulfill LTL specifications is an exceedingly formidable challenge (if not impossible), primarily due to uncountable nature of states and actions in continuous spaces.

To tackle the computational complexity challenges that arise, one promising solution is to approximate original (\emph{a.k.a.} concrete) systems with simpler models featuring finite state sets, commonly referred to as \emph{finite abstractions}~\cite{tabuada2009verification,belta2017formal}. When the underlying models are stochastic, these simplified representations commonly adopt the structure of Markov decision processes (MDPs), where discrete states mirror sets of continuous states in the concrete model (similarly for inputs). In practical implementation, constructing such finite abstractions usually entails partitioning the state and input sets of concrete models according to predefined discretization parameters, as discussed in various works including ~\cite{APLS08,julius2009approximations,zamani2014symbolic,tmka2013,hahn2013compositional,nejati2023formal,nejati2021compositional,lavaei2022scalable,lavaei2022automated}.

Within the finite MDP scheme, one can (i) initially leverage it as an appropriate substitute for the original system, (ii) proceed to synthesize controllers for the abstract system, and (iii) ultimately refine the controller back over the concrete model, facilitated by an interface map. Given that the disparity between the output of the original system and that of its abstraction is accurately quantified, it becomes feasible to ensure that the concrete system fulfills the same specification as the abstract counterpart under some quantified accuracy level. However, this accuracy level is only acceptable for finite-horizon specifications since it converges to infinity
as time approaches infinity (cf.~\eqref{eq:MDPaccuracy}), \emph{rendering MDPs impractical} for infinite-horizon properties.

As a promising alternative, \emph{interval Markov decision processes} (IMDPs)~\cite{sen2006model,GIVAN2000bounded} have emerged in the literature as a potential solution for formal verification and controller synthesis of stochastic systems fulfilling \emph{infinite-horizon specifications}. Specifically, IMDPs offer a comprehensive approach by incorporating both \emph{upper and lower} bounds on the transition probabilities among finite abstraction cells. This is accomplished by solving a (multi-player) game which allows to extend satisfaction guarantees to infinite time horizons. However, the construction of IMDPs is more complicated compared to traditional MDPs as it necessitates the computation of both lower and upper bounds for transition probabilities among partition cells.

Abstraction-based techniques, including those used for constructing either MDPs or IMDPs, encounter a significant challenge known as the \emph{curse of dimensionality}. This phenomenon refers to the exponential growth in computational complexity as the number of state dimensions increases. To alleviate this, we offer scalable parallel algorithms and efficient data structures designed for constructing IMCs/IMDPs and automating the process of verification and controller synthesis over infinite time horizons. In particular, by dividing the computations into smaller concurrent operations, we effectively mitigate the overall complexity by a factor equivalent to the number of threads available. This approach not only enhances computational efficiency of IMC/IMDP construction, but also facilitates the practical application of these techniques in real-world scenarios with high-dimensional spaces.

\subsection{Original Contributions}
The primary contributions and noteworthy aspects of our tool paper include:
\begin{enumerate}[(i)]
	\item We propose the first tool that constructs IMCs/IMDPs for large-scale discrete-time stochastic systems while providing \emph{convergence guarantees}. Our tool leverages the constructed IMCs/IMDPs for formal verification and controller synthesis ensuring the fulfillment of desired \emph{infinite-horizon} temporal logic specifications, encompassing \emph{safety, reachability, and reach-while-avoid properties}.
	\item \textsf{IMPaCT} is implemented in modern ISO C++ and runs in parallel using AdaptiveCpp\footnote{\href{https://github.com/AdaptiveCpp/AdaptiveCpp/}{https://github.com/AdaptiveCpp/AdaptiveCpp/}} based on SYCL\footnote{\href{https://www.khronos.org/sycl/}{https://www.khronos.org/sycl/}}. AdaptiveCpp eliminates from the user the need to implement cross-platform flexibility manually, serving as a strong foundation for CPU and GPU implementations.
	\item \textsf{IMPaCT} leverages \emph{interval iteration} algorithms to provide \emph{convergence guarantees} to an optimal controller in scenarios with \emph{infinite time horizons}.
	\item \textsf{IMPaCT} accepts bounded disturbances and natively supports additive noises with different \emph{practical distribution}s including normal and \emph{user-defined} distributions.
	\item We leverage \textsf{IMPaCT} across diverse real-world applications such as autonomous vehicles, room temperature systems, and building automation systems. This broadens the scope of formal method techniques to encompass safety-critical applications that require satisfaction within \emph{infinite time horizons}. The outcomes demonstrate significant efficiency in computational time.
\end{enumerate}

The source code for \textsf{IMPaCT} along with comprehensive instructions on how to build and operate it can be located at:
\begin{center}
	\href{https://github.com/Kiguli/IMPaCT}{https://github.com/Kiguli/IMPaCT}
	\vspace{-1mm}
\end{center}

\subsection{Related Literature}

IMDPs have become the source of significant interest inside the formal methods community over the past few years. In particular, IMDPs have been used for model-based verification and control problems~\cite{chen2013complexity,dutreix2020specification,hashemi2016compositional,weininger2019satisfiability,delimpaltadakis2023interval} as well as more recently being used for data-driven learning problems~\cite{jiang2022safe,lavaei2022constructing,rickard2023learning}.
There exists a limited set of software tools available for the formal verification and controller synthesis of stochastic systems, encompassing various classes of stochastic models, using abstraction-based techniques. \textsf{FAUST}$^{\mathsf 2}$ \cite{FAUST15} constructs finite MDPs for continuous-space discrete-time stochastic processes and conducts formal analysis for safety and reachability specifications. Nonetheless, the original MATLAB implementation of \textsf{FAUST}$^{\mathsf 2}$ encounters scalability issues, particularly for large models, due to the curse of dimensionality.

\textsf{StocHy}~\cite{StocHy} offers formal verification and synthesis frameworks for discrete-time stochastic hybrid systems via finite MDPs. While a small segment of \textsf{StocHy} addresses IMDPs, its primary implementation relies on the \emph{value iteration} algorithm~\cite{lahijanian2015formal}, which \emph{lacks convergence guarantees} when dealing with infinite-horizon specifications. This limitation has been widely pointed out in the relevant literature (see \emph{e.g.,}~\cite{haddad2014reachability,baier2017ensuring,haddad2018interval}). For the assurance of convergence to an optimal controller, it is essential to utilize \emph{interval iteration} algorithms. This is a key feature of our tool on top of offering parallelization,
which sets \textsf{IMPaCT} apart from \textsf{StocHy}, where value iteration is the default choice without any general parallelization capability. In particular, \textsf{IMPaCT} offers a notably more comprehensive and versatile approach for addressing \emph{infinite-horizon} specifications and stands as the first tool dedicated to IMC/IMDP construction with \emph{convergence guarantees}.

A recent update to \textsf{PRISM}~\cite{PRISM} supports \emph{robust} verification of uncertain models including IMDPs. However, \textsf{PRISM} requires the states and transitions of the IMDP to be described explicitly, which is cumbersome when the state and input spaces are large.
Furthermore, our tool introduces parallel implementations of IMDP abstraction and synthesis algorithms, efficiently automating the IMDP construction process. It is noteworthy to mention that \textsf{AMYTISS}~\cite{AMYTISS} also employs parallel implementations but exclusively for constructing MDPs using \textsf{PFaces}~\cite{PFACES}, built upon OpenCL, without any support for IMDP models. In addition, our tool is developed using SYCL, which facilitates parallel processing and offers broader utility than \textsf{PFaces}. In particular, AdaptiveCpp (formerly OpenSYCL)~\cite{SYCL1,SYCL2}, employed in \textsf{IMPaCT}, provides a sturdy foundation for flexible cross-platform parallel processing on CPUs, GPUs and hardware accelerators (HWAs) from all major hardware vendors such as Intel, NVIDIA, etc., which is not the case in \textsf{AMYTISS}. 

There exist a few more software tools for formal verification and controller synthesis of stochastic systems within diverse model classes, all without employing IMDP approaches. In this regard, \textsf{SReachTools}~\cite{vinod2019sreachtools} conducts stochastic reachability analysis specifically for linear (time-varying) discrete-time stochastic systems. \textsf{ProbReach}~\cite{shmarov2015probreach} is developed for the probabilistic reachability verification of stochastic hybrid systems. \textsf{SReach}~\cite{wang2015sreach} addresses 
probabilistic bounded reachability problems specifically within nonlinear hybrid automata affected by parametric uncertainty. \textsf{Modest Toolset}~\cite{hartmanns2014modest} enables modeling and analysis of distributed stochastic hybrid systems. \textsf{SySCoRe}~\cite{van2023syscore} is a MATLAB toolbox that derives simulation relations and presents an alternative methodology for the controller synthesis of stochastic systems satisfying co-safe specifications over infinite horizons. The last three ARCH workshops focused on friendly tool competitions over stochastic models are reported in \cite{abate2020arch,abate2022arch,abate2023arch}.

\subsection{Paper Organization}

This paper is structured as follows. In Section~\ref{sec:dt-SCS&MDP}, we define discrete-time stochastic control systems and their equivalent representations as continuous-space MDPs. In Section~\ref{sec:IMDP}, we define IMDPs, while presenting a serial algorithm for IMDP construction. Section~\ref{sec:IMDPAbstractionParallel} is dedicated to the parallel algorithms of IMDP construction. In Section~\ref{sec:IMDPSynthesisSerial}, we discuss serial algorithms for controller synthesis via constructed IMDPs, enforcing safety, reachability, and reach-while-avoid properties. Section~\ref{sec:IMDPSynthesisParallel} offers the parallel synthesis algorithms using constructed IMDPs. Section~\ref{sec:load&save} elaborates on the process of loading and saving respective files throughout various stages of the abstraction and synthesis procedure.
Section~\ref{sec:case_studies} showcases the outcomes of \textsf{IMPaCT} when implemented on well-known case studies and benchmark scenarios.

\subsection{Preliminaries}
We consistently employ the following notation throughout this work. The set of real numbers is denoted by $\reals$ and the set of natural numbers including zero by $\nats$. The empty set is denoted by $\emptyset$. We denote the cardinality of a set $A$ as $\vert A\vert$ and the power set as $2^A$. The Euclidean norm is denoted by $\lVert \cdot \rVert{_2}$, while the infinity norm is represented by $\lVert \cdot \rVert_{\infty}$. We use the notation $\mathsf{diag}(a_1, . . . , a_n)$ to represent a diagonal matrix in $\mathbb{R}^{n\times n}$, with its diagonal entries $a_1, . . . , a_n$ placed from the upper left corner. We consider a moment in time $k$ to be from the time horizon belonging to $\nats$. Where it is clear in context, time will be omitted for simplicity, \emph{e.g.} $x(k) \rightarrow x$. We consider a probability space ($\Omega, \mathcal{F}_\Omega, \mathbb{P}_\Omega$), where $\Omega$ is the sample space, $\mathcal{F}_\Omega$ is the sigma-algebra of $\Omega$ with elements of $\Omega$ called events, and $\mathbb{P}_\Omega$ is the probability measure that assigns a probability to each event. A topological space $X$ is called a Borel space if there exists a metric on $X$ that makes it a separable and complete metrizable space; $X$ is then endowed with a Borel sigma-algebra $\mathscr{B}(X)$. The measurable space of $X$ is ($X,\mathscr{B}(X)$). Other notation will be introduced when required.

\section{Discrete-Time Stochastic Control Systems}\label{sec:dt-SCS&MDP}

A formal description of discrete-time stochastic control systems, serving as the underlying dynamics of our tool, is presented in the following definition.
\begin{definition}[dt-SCS]\label{def:dt-SCS}
	A discrete-time stochastic control system (dt-SCS) is a quintuple
	\begin{equation}
	\label{eq:dt-SCS}
	\Sigma = (X,U,W,\varsigma,f),
	\end{equation}
	where,
	\begin{itemize}
		\item $X\subseteq \reals^n$ is a Borel space as the state set, with $(X,\borel(X))$ being its measurable space;
		\item $U\subseteq\reals^m$ is a Borel space as the input set;
		\item $W\subseteq\reals^p$ is a Borel space as the disturbance set;
		\item $\varsigma$ is a sequence of independent and identically distributed (i.i.d.) random variables from a sample space $\Omega$ to a measurable set $\mathcal{V}_\varsigma$ \[\varsigma:= \{\varsigma(k)\!:\Omega \rightarrow \mathcal{V}_\varsigma,~k\in\nats\};\]
		\item $f\!:X\times U\times W \times \mathcal{V}_\varsigma\rightarrow X$ is a measurable function characterizing the state evolution of the system.
	\end{itemize}
\end{definition}

For a given initial state $x(0)\in X$, an input sequence $u(\cdot):\Omega \rightarrow U$, and a disturbance sequence $w(\cdot):\Omega \rightarrow W$, the state evolution of $\Sigma$ is characterized by
\begin{equation}\label{eq:dt-SCS:dynamics}
\Sigma\!:x(k+1) = f(x(k),u(k),w(k),\varsigma(k)), \quad k\in\nats.
\end{equation}
To facilitate a more straightforward presentation of our contribution, we illustrate our algorithms by incorporating stochasticity with normal distributions. Nevertheless, our tool is capable of handling problems involving \emph{any arbitrary distributions} via a custom \emph{user-defined} distribution.

A dt-SCS has been shown to be equivalent to a continuous-space Markov decision process~\cite{kallenberg2021foundations} as the following definition, where a conditional stochastic kernel $T$ captures the evolution of $\Sigma$ and can be uniquely determined by the pair $(\varsigma, f)$ from~\eqref{eq:dt-SCS}.
\begin{definition}[Continuous-Space MDPs]
	\label{def:MDP}
	A continuous-space Markov decision process (MDP) is a quadtuple
	\begin{equation}
	\label{eq:MDP}
	\Sigma = (X,U,W,T),
	\end{equation}
	where, \begin{itemize}
		\item $X$, $U$, and $W$ are as in Definition~\ref{def:dt-SCS};
		\item $T\!:\borel(X) \times X \times U \times W \rightarrow [0,1]$ is a conditional stochastic kernel that assigns any $x\in X$, $u\in U$, and $w\in W$, a probability measure $T(\cdot\vert x,u,w)$, on the measurable space $(X,\borel(X))$ so that for any set $A\in\borel(X)$,
		\begin{equation*}
		\mathbb{P}\Big\{x(k+1)\in A \,\big\vert\, x(k), u(k), w(k)\Big\} = \int_A T(\text{d}x(k+1) \,\big\vert\, x(k), u(k), w(k)).
		\end{equation*}
	\end{itemize}
\end{definition}

To construct a traditional finite MDP (\emph{i.e.}, an MDP with finite spaces), the continuous spaces $X, U, W$ are constructed from a finite number of partitions $\textbf{X}^i, \textbf{U}^i, \textbf{W}^i$, where $X = \cup_{i=0}^{n_u}\textbf{X}^i, U = \cup_{i=0}^{n_x}\textbf{U}^i, W = \cup_{i=0}^{n_w}\textbf{W}^i$. The number of partitions $n_x, n_u, n_w$ can be computed based on discretization parameters $\eta_x, \eta_u, \eta_w$ which defines the size of regions $\textbf{X}^i, \textbf{U}^i, \textbf{W}^i$, respectively. Each finite partition can be identified by some representative points $\hat{x}_i\in\textbf{X}^i$, $\hat{u}_i\in\textbf{U}^i$, $\hat{w}_i\in\textbf{W}^i$;
the collections of all representative points can now be considered as the finite sets $\hat{X}, \hat{U}, \hat{W}$ of the finite MDP. Using a map $\Xi:X\rightarrow 2^X$, one can assign any continuous state $x\in X$ to the partition $\textbf{X}^i$, where $x\in\textbf{X}^i$. Additionally, a map $\Pi_x: X\rightarrow \hat{X}$ assigns any continuous state $x\in X$ to the representative point $\hat{x}\in\hat{X}$ of the corresponding partition containing $x$. The map $\Pi_x$ satisfies the inequality
\begin{equation*}
\lVert \Pi_x(x) - x\rVert_{2} \leq \eta_x, \quad\quad \forall x\in X,
\end{equation*}
with $\eta_x$ being a state discretization parameter. Using these mappings, the transition function $\hat{f}$ and the transition probability matrix $\hat{T}$ within the finite MDP construction are defined as
$
\hat{f}(\hat{x},\hat{u},\hat{w},\varsigma) = \Pi_x(f(\hat{x},\hat{u},\hat{w},\varsigma))$
and 
$
\hat{T}(\hat{x}'\vert \hat{x}, \hat{u}, \hat{w}) = T(\Xi(\hat{x}')\vert \hat{x}, \hat{u}, \hat{w})
$, respectively~\cite[Alg.~1 \& Thm.~2.2]{lavaei2017HSCC}.

Of particular importance to this work, an accuracy level $\rho$ regards the difference between probabilities of satisfaction of the desired specification $\psi$ over the continuous-space system $\Sigma$ and its finite MDP counterpart $\hat{\Sigma}$. This accuracy level $\rho$ can be computed a-priori as the product of the Lipschitz constant $\mathcal H$ of the stochastic kernel, the Lebesgue measure $\mathscr{L}$ of the state space, the state discretization parameter $\eta_x$, and the finite horizon $\mathcal K$, as the following:
\begin{equation}\label{eq:MDPaccuracy}
\big\vert \mathbb{P}(\Sigma \models \psi)- \mathbb{P}(\hat{\Sigma} \models \psi)\big\vert \leq \rho = \mathcal K \mathcal H\mathscr{L}\eta_x.
\end{equation}

\section{Interval Markov Decision Processes}\label{sec:IMDP}
The finite MDP presented in Section~\ref{sec:dt-SCS&MDP} is not suitable for infinite-horizon problems due to the inclusion of the time horizon $\mathcal K$ in \eqref{eq:MDPaccuracy}. In particular, as $\mathcal K \rightarrow \infty$ then $\rho \rightarrow \infty$, the distance between probabilities of satisfaction over concrete system $\Sigma$ and its finite MDP $\hat{\Sigma}$ converges to infinity, implying that the abstraction will be of no use to providing satisfaction guarantees. To alleviate this, a continuous-space MDP $\Sigma$ in~\eqref{eq:MDP} can be \emph{finitely abstracted} by an interval Markov decision process. Specifically, IMDPs provide bounds over the transition probability of the stochastic kernel $T$, offering a reliable model for analyzing infinite-horizon specifications, as outlined in the subsequent definition.

\begin{definition}[IMDPs]
	\label{def:IMDP}
	An interval Markov decision process (IMDP) is defined as a quintuple
	\begin{equation*}
	\hat{\Sigma} = (\hat{X},\hat{U},\hat{W},\hat{T}_{{\min}},\hat{T}_{{\max}}),
	\end{equation*}
	where,
	\begin{itemize}
		\item $\hat{X} = \{\hat{x}_0,\hat{x}_1,\ldots\hat{x}_{n_x}\}$, with $\hat{x}_i$ being the representative point within $\textbf{X}^i$, where $X = \cup_{i=0}^{n_x}\textbf{X}^i$;
		\item $\hat{U} = \{\hat{u}_0,\hat{u}_1,\ldots\hat{u}_{n_u}\}$, with $\hat{u}_i$ being the representative point within $\textbf{U}^i$, where $U = \cup_{i=0}^{n_u}\textbf{U}^i$;
		\item $\hat{W} = \{\hat{w}_0,\hat{w}_1,\ldots\hat{w}_{n_w}\}$, with $\hat{w}_i$ being the representative point within $\textbf{W}^i$, where $W = \cup_{i=0}^{n_w}\textbf{W}^i$;
		\item $\hat{T}_{{\min}}$
		is a conditional stochastic kernel for the minimal transition probability, computed as
		\begin{align*}
		\hat{T}_{{\min}}(\hat{x}'\,\vert\, \hat{x},\hat{u},\hat{w}) = \min\limits_{x\in\Xi(\hat{x})}~T(\Xi(\hat{x}')\,\vert\, x,\hat{u},\hat{w}),\quad
		\forall\hat{x},\hat{x}'\in \hat{X},~\forall\hat{u}\in\hat{U},~\forall\hat{w}\in\hat{W},
		\end{align*}
		with $x\in \Xi(\hat{x})$, and where the map $\Xi:X\rightarrow 2^{X}$ assigns to any $x\in X$, the corresponding partition element it belongs to, i.e., $\Xi(x) = \textbf{X}^i$ if $x\in\textbf{X}^i$;
		\item $\hat{T}_{{\max}}$ is a conditional stochastic kernel for the maximal transition probability, computed similarly as
		\begin{align*}
		&\hat{T}_{{\max}}(\hat{x}'\,\vert\, \hat{x},\hat{u},\hat{w}) = \max\limits_{x\in\Xi(\hat{x})}~T(\Xi(\hat{x}')\,\vert\, x,\hat{u},\hat{w}),\quad
		\forall\hat{x},\hat{x}'\in \hat{X},~\forall\hat{u}\in\hat{U},~\forall\hat{w}\in\hat{W},\\
		&\hat{T}_{{\min}}\leq\hat{T}_{{\max}}, ~\text{and}~ \sum\limits_{\hat x'\in \hat{X}}\hat{T}_{{\min}}(\hat{x}'\,\vert\, \hat{x},\hat{u},\hat{w}) \leq 1 \leq \sum\limits_{\hat x'\in \hat{X}}\hat{T}_{{\max}}(\hat{x}'\,\vert\, \hat{x},\hat{u},\hat{w}).
		\end{align*}
	\end{itemize}
\end{definition}

Although IMDPs  incur higher computational costs compared to traditional MDPs, leveraging lower and upper bound probabilities for state transitions facilitates the resolution of infinite-horizon control problems. It is worth mentioning that since $\hat X,\hat U,\hat W$ are all finite sets, $\hat{T}_{{\min}}$ and $\hat{T}_{{\max}}$ can be represented by static matrices of size $(n_x \times n_u \times n_w)$ by $n_x$.
This enables the use of powerful iterative algorithms.

\subsection{Temporal Logic Specifications}\label{new}

Here, we define the specifications of interest handled by \textsf{IMPaCT}. The desired specification $\psi$ is codified using \emph{linear temporal logic} (LTL)~\cite{katoen2008book}. Of particular interest are the properties described via logical operators including \emph{always} $\square$,
\emph{eventually} $\lozenge$, and
\emph{until} $\until$. 
\begin{definition}[Specifications]
	The specifications of interest  $\psi$, handled by \textsf{IMPaCT}, are defined as
	\begin{itemize}
		\item $\psi:=\square{\mathcal{S}}$ -
		safety; the system should always remain within a safe region ${\mathcal{S}}\subseteq X$.
		\item $\psi:=\lozenge\mathcal{T}$ - reachability; the system should eventually reach some target region $\mathcal{T}\subseteq X$.
		\item $\psi:={\mathcal{S}} \until \mathcal{T}$ - reach-while-avoid;
		the system should remain within the safe region ${\mathcal{S}}\subseteq X\backslash \mathcal{A}$ until it reaches the target region $\mathcal{T}\subseteq X$, with $\mathcal{A}$ being an avoid region. 
	\end{itemize}
\end{definition}

When constructing IMCs/IMDPs, we aim at labeling the states within the state space based on the specification, especially beneficial for later-stage verification or controller synthesis processes. To do so, with a slight abuse of notation, we consider states $\hat{x}\in\mathcal{S}$ to refer explicitly to states in the state space considered to be safe, \emph{i.e.,} for safety specifications these states belong to the safe region $\mathcal{S}$. We now relabel states from the state space that belong to the target set as target states $\hat{r}\in\mathcal{T}$ and assume that all of these states are absorbing states, \emph{i.e.,} if the system reaches the target region, it will remain within that region. Similarly, we relabel any states from the state space that also belong to the avoid region as avoid states $\hat{a}\in\mathcal{A}$, treating the avoid region as an absorbing area. Due to the absorbing properties, the avoid and target regions are commonly modeled as a single state to simplify the algorithms. Transition probabilities to these states can be summed together for these new states. According to this implementation technique, we introduce static vectors that capture the minimum and maximum probabilities of transitions to the target region, with row entries calculated by
\begin{align*}
&\hat{R}_{\min}(\hat{x},\hat{u},\hat{w}) = \sum\limits_{\forall\hat{r}\in\mathcal{T}}\hat{T}_{\min}(\hat{r}\,\vert\,\hat{x},\hat{u},\hat{w}), \quad \hat{R}_{\max}(\hat{x},\hat{u},\hat{w}) = \sum\limits_{\forall\hat{r}\in\mathcal{T}}\hat{T}_{\max}(\hat{r}\,\vert\,\hat{x},\hat{u},\hat{w}).
\end{align*}
Similarly, for transitions to the avoid region, we define 
\begin{align*}
\hat{A}_{\min}(\hat{x},\hat{u},\hat{w}) = \sum\limits_{\forall\hat{a}\in\mathcal{A}}\hat{T}_{\min}(\hat{a}\,\vert\,\hat{x},\hat{u},\hat{w}), \quad \hat{A}_{\max}(\hat{x},\hat{u},\hat{w})= \sum\limits_{\forall\hat{a}\in\mathcal{A}}\hat{T}_{\max}(\hat{a}\,\vert\,\hat{x},\hat{u},\hat{w}).
\end{align*}
Therefore, one can redefine the IMDP depending on the specification as $\hat{\Sigma} = (\hat X, \hat U, \hat W, \hat{T}_{{\min}},\hat{T}_{{\max}},\hat{R}_{{\min}},\hat{R}_{{\max}},\\\hat{A}_{{\min}},\hat{A}_{{\max}})$. It is worth noting that users do not need to be aware of these additional vectors, as they are automatically computed during the abstraction process from the state labeling, as detailed in Algorithm~\ref{alg:abstractionserial}.

\subsection{Construction of IMDP}
\begin{algorithm}[h]
	\caption{Serial construction of IMDP}
	\label{alg:abstractionserial}
	\SetAlgoLined
	\KwData{Continious MDP $\Sigma = (X,U,W,T)$ and specification $\psi$}
	Select finite partition of sets $X, U, W$ as       
	$X =\cup^{n_x}_{i=0}\textbf{X}^i$, $U = \cup^{n_u}_{i=0}\textbf{U}^i$, $W = \cup^{n_w}_{i=0}\textbf{W}^i$\;
	
	For each $\textbf{X}^i$ select
	a representative point $\hat{x}^i\in\textbf{X}^i$. Similarly,
	for each $\textbf{U}^i$ and $\textbf{W}^i$, select 
	a representative point $\hat{u}^i\in\textbf{U}^i$ and $\hat{w}^i\in\textbf{W}^i$\;
	Define $\hat{X} := \{\hat{x}^i, i = 0,\ldots,n_x\}$ as the finite state set of $\hat\Sigma$ with input set $\hat{U} := \{\hat{u}^i, i = 0,\ldots,n_u\}$ and disturbance set $\hat{W} := \{\hat{w}^i, i = 0,\ldots,n_w\}$\;
	Define the map $\Xi: X \rightarrow 2^X$ that assigns to any $x\in X$, the corresponding partition set it belongs to, \emph{i.e.}, $\Xi(x) = \textbf{X}^i$ if $x \in \textbf{X}^i$\;
	Label states as safe states $\hat{x}\in\mathcal{S}$, target states $\hat{r}\in\mathcal{T}$, or avoid states $\hat{a}\in\mathcal{A}$, based on desired specification $\psi$\;
	\For{$\hat{w}_i\in\hat{W},i=\{0,\ldots,n_w\}$}{
		\For{$\hat{u}_j\in\hat{U},j=\{0,\ldots,n_u\}$}{
			\For{$\hat{x}_k\in\mathcal{S},k=\{0,\ldots,{n_s}\}$
			}{
				\For{$\hat{x}'_l\in\mathcal{S},l=\{0,\ldots,{n_s}\}$
				}{
					$\hat{T}_{{\min}}(\hat{x}'_l\,\vert\, \hat{x}_k,\hat{u}_j,\hat{w}_i) = \min\limits_{x\in\Xi(\hat{x}_k)}T(\Xi(\hat{x}'_l)\,\vert\, x,\hat{u}_j,\hat{w}_i)$\; $\hat{T}_{{\max}}(\hat{x}'_l\,\vert\, \hat{x}_k,\hat{u}_j,\hat{w}_i) = \max\limits_{x\in\Xi(\hat{x}_k)}T(\Xi(\hat{x}'_l)\,\vert\, x,\hat{u}_j,\hat{w}_i)$\;}
				{$\hat{R}_{{\min}}(\hat{x}_k,\hat{u}_j,\hat{w}_i) = \sum\limits_{\forall\hat{r}\in\mathcal{T}}\min\limits_{x\in\Xi(\hat{x}_k)}T(\Xi(\hat{r})\,\vert\, x,\hat{u}_j,\hat{w}_i)$\; $\hat{R}_{{\max}}(\hat{x}_k,\hat{u}_j,\hat{w}_i) = \sum\limits_{\forall\hat{r}\in\mathcal{T}}\max\limits_{x\in\Xi(\hat{x}_k)}T(\Xi(\hat{r})\,\vert\, x,\hat{u}_j,\hat{w}_i)$\;
					$\hat{A}_{{\min}}(\hat{x}_k,\hat{u}_j,\hat{w}_i) = \sum\limits_{\forall\hat{a}\in\mathcal{A}}\min\limits_{x\in\Xi(\hat{x}_k)}T(\Xi(\hat{a})\,\vert\, x,\hat{u}_j,\hat{w}_i)$\; $\hat{A}_{{\max}}(\hat{x}_k,\hat{u}_j,\hat{w}_i) = \sum\limits_{\forall\hat{a}\in\mathcal{A}}\max\limits_{x\in\Xi(\hat{x}_k)}T(\Xi(\hat{a})\,\vert\, x,\hat{u}_j,\hat{w}_i)$\;}}}}
	\KwResult{{IMDP $\hat\Sigma = (\hat X, \hat U, \hat W, \hat{T}_{{\min}},\hat{T}_{{\max}},\hat{R}_{{\min}},\hat{R}_{{\max}},\hat{A}_{{\min}},\hat{A}_{{\max}})$}}
\end{algorithm}

We now describe the approach taken to construct an IMDP $\hat{\Sigma}$, including how to compute minimal and maximal transition probability matrices $\hat{T}_{{\min}}$, $\hat{T}_{{\max}}$, $\hat{R}_{{\min}}$, $\hat{R}_{{\max}}$, $\hat{A}_{{\min}}$, and $\hat{A}_{{\max}}$, which is essential for later-stage verification or controller synthesis procedures. We first present a serial algorithm for the construction of IMDPs, as Algorithm~\ref{alg:abstractionserial}, considering stochasticity with normal distributions which are the default noise type in \textsf{IMPaCT}, while providing a custom user-defined distribution to support \emph{any arbitrary distributions}.

In Algorithm~\ref{alg:abstractionserial}, Steps $1-2$
first construct finite partitions and obtain the representative points within $X$, $U$, and $W$.
In Step $3$,
the finite set of states $\hat{X}$, inputs $\hat{U}$, and disturbances $\hat{W}$ are defined, where $n_x = \vert \hat{X}\vert$, $n_u = \vert \hat{U}\vert$, and $n_w = \vert \hat{W}\vert$. In Step $5$, we filter the states based on their labeling for the specification $\psi$. In Steps $10 - 11$, the transition probability matrices $\hat{T}_{{\min}}$ and $\hat{T}_{{\max}}$ are computed by iterating through all the combinations of safe states $\hat{x}\in\mathcal{S}$, where $n_s = \vert\mathcal{S}\vert$, inputs $\hat{u}\in\hat{U}$, and disturbances $\hat{w}\in\hat{W}$. To do so, a nonlinear optimization is required to find $x\in\Xi(\hat{x})$ corresponding to either the lower bound or upper bound probability of transitioning from $\Xi(\hat{x})$ to $\Xi(\hat{x}')$. Similarly, Steps $13-16$ perform the same procedure for one-step transitions going to either the target or avoid regions.
The output of the algorithm is therefore IMDP $\hat\Sigma = (\hat X, \hat U, \hat W, \hat{T}_{{\min}},\hat{T}_{{\max}},\hat{R}_{{\min}},\hat{R}_{{\max}},\hat{A}_{{\min}},\hat{A}_{{\max}})$.

\begin{remark}
	One can readily construct interval Markov chains (IMCs) for verification purposes via Algorithm~\ref{alg:abstractionserial} by setting the control input to zero within the concrete dynamics, \emph{i.e.,} $\Sigma = (X,W,T)$. 
\end{remark}

\subsection{Complexity Analysis for Serial Construction of IMDP}
To conduct a computational complexity analysis for the serial construction of IMDP, we adopt a slight abuse of notation by defining $n_{x_i}$, $n_{u_j}$, and $n_{w_k}$ as the dimension-wise counts of partitions within $\hat{X}$, $\hat{U}$, and $\hat{W}$, respectively, where $i = 1,\ldots, n$, $j = 1,\ldots, m$, and $k = 1,\ldots, p$. The computational complexity of IMDP construction comprises two main components. The first, often termed the \emph{curse of dimensionality}, describes the exponential rise in computational time concerning system dimensions, expressed as $\mathcal{O}(2d)$, where $d = (n_{x_i}^n\times n_{u_j}^m \times n_{w_k}^p) \times n_{x_i}^n$. The value $d$ represents the total count of rows in the static transition matrix, \emph{i.e.,} $n_x \times n_x \times n_u$, multiplied by the number of columns, \emph{i.e.,} $n_x$. Given that the construction of IMDP abstraction involves both lower bound and upper bound transition matrices $\hat{T}_{\min}$ and $\hat{T}_{\max}$, respectively, the computational complexity $d$ is doubled.

The second part pertains to the computational complexity of the nonlinear optimization within Steps $10-11$ in Algorithm~\ref{alg:abstractionserial}, represented as $\mathcal{O}({\kappa})$. The level of complexity $\kappa$ hinges on the algorithm chosen by the user, detailed in Section~\ref{sec:nnlopt}. Consequently, the overall complexity of Algorithm~\ref{alg:abstractionserial} amounts to $\mathcal{O}({2\kappa}d)$. This accounts for both the space explosion ($d$) due to the system dimensions and the computational load ($\kappa$) associated with nonlinear optimization.

\section{Parallel Construction of IMDP}\label{sec:IMDPAbstractionParallel}
In this section, we describe how we update Algorithm~\ref{alg:abstractionserial} for parallel processing as offered by \textsf{IMPaCT}. We also here present the user functions of our tool corresponding to various subsections. Firstly, an IMDP object should be constructed by defining the dimensions of state, input and disturbance:

\begin{lstlisting}[language=C++,caption=Creating IMDP object.]
IMDP(const int x, const int u, const int w);
\end{lstlisting}

\subsection{Setting Noise Distribution}

\textsf{IMPaCT} by default supports normal distributions but also allows a user-defined distribution to be provided:

\begin{lstlisting}[language=C++,caption=Define noise distribution.]
enum class NoiseType {NORMAL, CUSTOM};
\end{lstlisting}

We consider here an i.i.d. noise where the tool receives a vector of standard deviations. In this setup, the covariance matrix is diagonal, and each diagonal entry represents the variance:

\begin{lstlisting}[language=C++,caption= {Diagonal covariance matrix}.]
void setStdDev(vec sig);
void setNoise(NoiseType n, bool diagonal = true);
\end{lstlisting}

\textsf{IMPaCT} also accommodates full covariance matrix, with nonzero off-diagonal elements, when provided with the inverse covariance matrix and its determinant. For multi-dimensional noise distributions, Monte Carlo integration is employed instead of a closed-form solution, offering advantages for higher dimension integrals~\cite{newman1999monte}, requiring the user to specify the number of samples for the integration process. We utilize the GNU Scientific Library for this integration process~\cite{gough2009gnu}. In a broader scope, matrix and vector manipulations are implemented using the C++ library Armadillo~\cite{sanderson2016armadillo,sanderson2018user}:

\begin{lstlisting}[language=C++,caption={Full covariance matrix}.]
void setInvCovDet(mat inv_cov, double det);
void setNoise(NoiseType n, bool diagonal, size_t monte_carlo_samples);
\end{lstlisting}

We facilitate the utilization of a user-defined custom distribution, prompting the user to define the specific desired distribution. The code will provide a struct \verb$customParams$ containing data for the variables \verb$state_start$ $\hat{x}$, \verb$input$ $\hat{u}$, \verb$disturb$ $\hat{w}$, \verb$eta$ $\eta_x$, \verb$mean$ $f(\hat{x},\hat{u},\hat{w})$, \verb$lb$ lower bound of the integration, \verb$ub$ upper bound of the integration, and \verb$dynamics{1,2,3}$ dynamics with numbers representing the number of function parameters. The \verb$customPDF$ function intended for integration can be coded within the file \verb$src/custom.cpp$, and the parameters for the Monte Carlo integration--\verb$CUSTOM$ noise and the number of samples--can be configured using:

\begin{lstlisting}[language=C++,caption=Custom noise distributions.]
void setNoise(NoiseType n);
void setCustomDistribution(size_t monte_carlo_samples);
\end{lstlisting}

\subsection{Parallel Construction of Transition Probability Matrices}

We propose in Algorithm~\ref{alg:abstraction} the parallel approach for constructing the transition matrices $\hat{T}_{{\min}}$, $\hat{T}_{{\max}}$, $\hat{R}_{{\min}}$, $\hat{R}_{{\max}}$, $\hat{A}_{{\min}}$, and $\hat{A}_{{\max}}$. Steps $1-5$ are equivalent to Algorithm~\ref{alg:abstractionserial}. In \textsf{IMPaCT}, these Steps $1-5$, can be computed using the following user functions. The dimensions and representative points within each space can be specified by setting the lower bounds, upper bounds, and discretization parameters. As elaborated in Subsection~\ref{new}, we consider states within the state space as safe states by default, constituting the safe region. Subsequently, the target and avoid regions can be delineated by employing Boolean expressions that assess the state space. These expressions facilitate the relabeling of states, transitioning them from safe states to either target or avoid states based on desired specifications:
\begin{lstlisting}[language=C++, caption={Construction of $\hat{X},\hat{U},\hat{W},\mathcal{T}$, and $\mathcal{A}$}.]
void setStateSpace(vec lb, vec ub, vec eta);
void setInputSpace(vec lb, vec ub, vec eta);
void setDisturbSpace(vec lb, vec ub, vec eta);
void setTargetSpace(const function<bool(const vec&)>& separate_condition, bool remove);
void setAvoidSpace(const function<bool(const vec&)>& separate_condition, bool remove);
void setTargetAvoidSpace(const function<bool(const vec&)>& target_condition,const function<bool(const vec&)>& avoid_condition, bool remove);
\end{lstlisting}

\begin{algorithm}[h]
	\caption{\emph{Parallel} Construction of IMDP}
	\label{alg:abstraction}
	\SetAlgoLined
	\KwData{MDP $\Sigma = (X,U,W,T)$ and specification $\psi$}
	Select finite partition of sets $X, U, W$ as:       
	$X =\cup^{n_x}_{i=0}\textbf{X}^i$, $U = \cup^{n_u}_{i=0}\textbf{U}^i$, $W = \cup^{n_w}_{i=0}\textbf{W}^i$\;
	
	For each partition $\textbf{X}^i$ select a representative point $\hat{x}^i\in\textbf{X}^i$ used to describe the partition, \emph{e.g.,} centre of the partition.
	Similarly for $\textbf{U}^i$ and $\textbf{W}^i$, select representative points $\hat{u}^i\in\textbf{U}^i$ and $\hat{w}^i\in\textbf{W}^i$\;
	Define $\hat{X} := \{\hat{x}^i, i = 0,\ldots,n_x\}$ as the finite state set of $\hat\Sigma$ with input set $\hat{U} := \{\hat{u}^i, i = 0,\ldots,n_u\}$ and disturbance set $\hat{W} := \{\hat{w}^i, i = 0,\ldots,n_w\}$\;
	Define the map $\Xi: X \rightarrow 2^X$ that assigns to any $x\in X$, the corresponding partition set it belongs to, \emph{i.e.}, $\Xi(x) = \textbf{X}^i$ if $x \in \textbf{X}^i$\;
	{Relabel states as safe states $\hat{x}\in\mathcal{S}$, target states $\hat{r}\in\mathcal{T}$, or avoid states $\hat{a}\in\mathcal{A}$ based on desired specification $\psi$\;}
	\ForAll{combinations of $\hat{x}_k\in\mathcal{S},\hat{u}_j\in\hat{U},$ and $\hat{w}_i\in\hat{W}$ \textbf{in parallel}}{ 
		{\For{$\hat{x}'_l\in\mathcal{S},l=\{0,\ldots,n_x\}$}{
				$\hat{T}_{{\min}}(\hat{x}'_l\vert \hat{x}_k,\hat{u}_j,\hat{w}_i) = \min\limits_{x\in\Xi(\hat{x}_k)}T(\Xi(\hat{x}'_l)\vert x,\hat{u}_j,\hat{w}_i)$\; $\hat{T}_{{\max}}(\hat{x}'_l\vert \hat{x}_k,\hat{u}_j,\hat{w}_i) = \max\limits_{x\in\Xi(\hat{x}_k)}T(\Xi(\hat{x}'_l)\vert x,\hat{u}_j,\hat{w}_i)$\;}
			{$\hat{R}_{{\min}}(\hat{x}_k,\hat{u}_j,\hat{w}_i) = \sum\limits_{\forall\hat{r}\in\mathcal{T}}\min\limits_{x\in\Xi(\hat{x}_k)}T(\Xi(\hat{r})\vert x,\hat{u}_j,\hat{w}_i)$\; $\hat{R}_{{\max}}(\hat{x}_k,\hat{u}_j,\hat{w}_i) = \sum\limits_{\forall\hat{r}\in\mathcal{T}}\max\limits_{x\in\Xi(\hat{x}_k)}T(\Xi(\hat{r})\vert x,\hat{u}_j,\hat{w}_i)$\;
				$\hat{A}_{{\min}}(\hat{x}_k,\hat{u}_j,\hat{w}_i) = \sum\limits_{\forall\hat{a}\in\mathcal{A}}\min\limits_{x\in\Xi(\hat{x}_k)}T(\Xi(\hat{a})\vert x,\hat{u}_j,\hat{w}_i)$\; $\hat{A}_{{\max}}(\hat{x}_k,\hat{u}_j,\hat{w}_i) = \sum\limits_{\forall\hat{a}\in\mathcal{A}}\max\limits_{x\in\Xi(\hat{x}_k)}T(\Xi(\hat{a})\vert x,\hat{u}_j,\hat{w}_i)$\;}}}
	\KwResult{IMDP $\hat\Sigma = (\hat X, \hat U, \hat W, \hat{T}_{{\min}},\hat{T}_{{\max}},\hat{R}_{{\min}},\hat{R}_{{\max}},\hat{A}_{{\min}},\hat{A}_{{\max}})$}
\end{algorithm}

In Steps $8-9$, transition probabilities are computed \emph{in parallel} for the value $x\in\Xi(\hat{x})$ corresponding to either the minimal (lower bound) or maximal (upper bound) probability of transitioning from $\Xi(\hat{x})$ to $\Xi(\hat{x}')$. In Step $11-12$, target vector bounds $\hat{R}_{{\min}}$ and $\hat{R}_{{\max}}$ calculate the one-step transition probability of the state $\hat{x}$ transitioning to $\mathcal{T}$. Similarly, avoid vector bounds $\hat{A}_{{\min}}$ and $\hat{A}_{{\max}}$ calculate the transition probability of the state $\hat{x}$ transitioning to $\mathcal{A}$. For each state-input-disturbance triple, we sum the probabilities together resulting in the respective target vector or avoid vector. In \textsf{IMPaCT}, Steps $8-9$ and $11-14$, can be computed using the following function calls:

\begin{lstlisting}[language=C++,caption=Abstraction of transition matrices.]
void minTransitionMatrix();
void maxTransitionMatrix();
void minTargetTransitionVector();
void maxTargetTransitionVector();
void minAvoidTransitionVector();
void maxAvoidTransitionVector();
\end{lstlisting}

\begin{remark}
	In \textsf{IMPaCT}, given that the state space is bounded, some case studies may involve potential transitions outside the defined state space. This is captured in the functions \verb$minAvoidTransitionVector()$ as $\hat{A}_{\min}(\hat{x},\hat{u},\hat{w}) = 1 - \hat{T}_{\max}(\hat{X}\vert \hat{x},\hat{u},\hat{w})$ and with a similar construction for \verb$maxAvoidTransitionVector()$. These functions also consider the minimal and maximal transition probabilities to any additional avoid states that are inside the state space by summing them to $\hat{A}_{\min}(\hat{x},\hat{u},\hat{w})$ and $\hat{A}_{\max}(\hat{x},\hat{u},\hat{w})$, respectively.
\end{remark}

\subsection{Complexity Analysis for Parallel Construction of IMDP}
\label{sec:nnlopt}
Running these steps in parallel enhances the performance of solving the abstraction and synthesis problems. Using $\mathcal{O}(\kappa)$ to represent the optimization complexity, the complexity of Algorithm~\ref{alg:abstraction} is therefore $\mathcal{O}(\frac{2\kappa d}{\textsc{threads}}),$ where $d = (n_{x_i}^n\times n_{u_j}^m \times n_{w_k}^p) \times n_{x_i}^n$, and $\textsc{threads}$ is the number of parallel threads running. We implement the optimization algorithms using NLopt~\cite{NLopt}, by default we select \verb$nlopt::LN_SBPLX$~\cite{SUBPLEX}, which is a variant of the derivative-free Nelder-Mead Simplex algorithm.
It is worth highlighting that the chosen algorithm in \textsf{IMPaCT} can be readily swapped to any other nonlinear optimization algorithm from NLopt\footnote{\href{https://nlopt.readthedocs.io/en/latest/NLopt_Algorithms/}{https://nlopt.readthedocs.io/en/latest/NLopt$\_$Algorithms/}} using the following function:

\begin{lstlisting}[language=C++,caption=Set NLopt algorithm.]
void setAlgorithm(nlopt::algorithm alg);
\end{lstlisting}

\subsection{Low-cost abstraction} The primary bottleneck in the IMDP abstraction arises from the necessity of computing two transition probability matrices, coupled with the additional computational load of $\min$/$\max$ optimization. To enhance performance, by computing the matrix $\hat{T}_{{\max}}$ first, the computations required for $\hat{T}_{{\min}}$ can be reduced. This is due to the fact that for any matrix entry $\hat{T}_{{\max}}(i,j) = 0$, one has $\hat{T}_{{\min}}(i,j) = 0$. The sparser the matrix, the more efficient the abstraction computation; we refer to this as the \emph{low-cost abstraction}. This speed up can be used by calling the following functions for the transition matrices and the target transition vectors, respectively:

\begin{lstlisting}[language=C++,caption=Low-cost abstractions.]
void transitionMatrixBounds();
void targetTransitionMatrixBounds();
\end{lstlisting}

\section{Controller Synthesis with Interval Iteration}
\label{sec:IMDPSynthesisSerial}

We now synthesize a controller, via constructed IMDP, to enforce infinite-horizon properties over dt-SCS.
When dealing with infinite-horizon problems, the \emph{interval iteration} algorithm is capable of providing convergence guarantees unlike the common \emph{value iteration}~\cite{haddad2018interval}. In particular, the interval iteration algorithm converges to an under-approximation and over-approximation of the satisfaction probability associated with a temporal logic specification. The trade-off for achieving such a guarantee is the doubled computational load compared to the value iteration algorithm.

In essence, the interval iteration algorithm iterates over two Bellman equations simultaneously; one assuming an initial probability vector of zeros, denoted by $V_0 = \textbf{0}_n$, and the other an initial vector of ones, represented as $V_1 = \textbf{1}_n$. When calculating $V_0'$ and $V_1'$ as the next iteration step, a dynamic program with decision variables $\hat{R}$, $\hat{A}$, and $\hat{T}$ needs to be solved~\cite{haddad2018interval}. In finite MDP construction, a state within a partition is predetermined beforehand, and the probability transitions to all other partitions are computed based on this state. However, in IMDP construction, each partition might utilize a distinct state when calculating probability transitions to another partition, contingent upon the destination partition, \emph{i.e.,} the partition where the system lands after a one-step evolution. Consequently, solving the dynamic program can be seen as akin to determining a partition's state that could universally serve for computing probability transitions to other partitions. This solution is referred to as a \emph{feasible} distribution. In particular, the feasible distribution is determined to optimize (either minimize or maximize) the decision variables given the desired property of interest. Subsequently, the corresponding optimization is minimized with respect to the disturbance $\hat{w}$ and maximized concerning the input $\hat{u}$:
\begin{equation}
\label{eq:linear-program}    \max\limits_{\hat{u}\in\hat{U}}\min\limits_{\hat{w}\in\hat{W}} \underset{\hat{R},\hat{A},\hat{T}}{\text{optimize}}~\delta_1\hat{R}(\hat{x},\hat{u},\hat{w}) + \delta_2\hat{A}(\hat{x},\hat{u},\hat{w}) + \sum\limits_{\forall\hat{x}'\in\hat{X}}\delta_3(\hat{x}')\hat{T}(\hat{x}'\vert\hat{x},\hat{u},\hat{w})
\end{equation}
with weight functions $\delta_1$, $\delta_2$, and $\delta_3(\cdot)$, and being subject to the following constraints~\cite{haddad2018interval}:
\begin{align*}
&\hat{T}_{\min}(\hat{x}'\vert\hat{x},\hat{u},\hat{w})\leq \hat{T}(\hat{x}'\vert\hat{x},\hat{u},\hat{w})\leq\hat{T}_{\max}(\hat{x}'\vert\hat{x},\hat{u},\hat{w}), \\ 
&\hat{R}_{\min}(\hat{x},\hat{u},\hat{w})\leq \hat{R}(\hat{x},\hat{u},\hat{w})\leq\hat{R}_{\max}(\hat{x},\hat{u},\hat{w}),\\
&\hat{A}_{\min}(\hat{x},\hat{u},\hat{w})\leq \hat{A}(\hat{x},\hat{u},\hat{w})\leq\hat{A}_{\max}(\hat{x},\hat{u},\hat{w}),  \\ &\hat{R}(\hat{x},\hat{u},\hat{w})+\hat{A}(\hat{x},\hat{u},\hat{w}) +\sum\limits_{\forall\hat{x}'\in\hat{X}}\hat{T}(\hat{x}'\vert\hat{x},\hat{u},\hat{w}) = 1.
\end{align*} 
In particular, $\hat{R}$ mimics $s^+$ and $\hat{A}$ mimics $s^-$ in the formulation of~\cite{haddad2018interval}. The corresponding weights are attributed to the dynamic program and are updated at each iteration step. The controller synthesis here uses a linear program (LP) to converge to an optimal solution, where the weights are the probability of satisfying the specification and $\delta_3$ is derived from either $V_0$ or $V_1$. Specifically, the optimization in~\eqref{eq:linear-program} takes into account the probability from transitioning from a partition with representative state $\hat{x}$ to another partition with representative state $\hat{x}'$ multiplied by some weight. This weight corresponds to the probability of satisfying the specification when initiating from a state within the partition indicated by $\hat{x}'$. The affine part of the optimization depends on the specification for the weights attributed to $\hat{R}$ and $\hat{A}$. For reachability specifications, $\delta_1 = 1$ and $\delta_2 = 0$, whereas for safety, the inverse holds true. Additionally, the probability of fulfilling the specification complements the resulting optimal values for $V_0$ and $V_1$.

When the dynamic program is solved and the optimal feasible solutions $\hat{T}$, $\hat{R}$, and $\hat{A}$ are found, the interval iteration algorithm solves the two equations
\begin{equation}
\begin{cases}
V_0' = \delta_1\hat{R} + \delta_2\hat{A} + \hat{T}V_0,\\
V_1' = \delta_1\hat{R} + \delta_2\hat{A} + \hat{T}V_1,
\end{cases}
\end{equation}
to find the new probabilities of satisfying the specification for the state-input-disturbance triples. Prior to the subsequent iteration, $V_0$ and $V_1$ are, respectively, updated to $V_0'$ and $V_1'$. The interval iteration algorithm terminates when the two vectors converge, $\lVert V_1 - V_0\rVert_{\infty} \leq \varepsilon$,
where $\varepsilon$ is set by default to a small enough threshold.

The resulting controller $\mathcal{C} = (\hat{X},\pi,\mathds {P}_{\psi_{\min}},\mathds {P}_{\psi_{\max}})$
consists of a lookup table with the optimal control policy $\pi$ for each representative point $\hat{x}\in\mathcal{S}$, the minimal probability of satisfying the specification
$\mathds {P}_{\psi_{\min}}$
to and the maximal probability of satisfying the specification
$\mathds {P}_{\psi_{\max}}$. When synthesizing the controller, either $\mathds {P}_{\psi_{\min}}$ or $\mathds {P}_{\psi_{\max}}$ is prioritized for finding $\pi$, giving a \emph{pessimistic or optimistic} control policy, respectively. In particular, a pessimistic policy is when the policy $\pi$ is optimized using~\eqref{eq:linear-program} that minimizes $\hat{A},\hat{R}$ and $\hat{T}$, while an optimistic policy maximizes $\hat{A},\hat{R}$ and $\hat{T}$~\cite{delimpaltadakis2023interval}. The policy $\pi$ is then fixed when calculating the other bound, \emph{e.g.,} $\pi$ synthesized from calculations on $\mathds {P}_{\psi_{\min}}$ is fixed when finding $\mathds {P}_{\psi_{\max}}$.

The underlying synthesis technique is equivalent to a three-and-a-half player game with a max-min optimization problem, where the input and disturbance are two players. In particular, treating the disturbance as an adversary to the system entails minimizing the probability concerning the disturbance while maximizing it concerning the control inputs. The third player represents the range across the targeted partition addressed by the optimization program, acting as an adversary in computing $\mathds{P}_{\psi_{\min}}$ and an ally when computing $\mathds{P}_{\psi_{\max}}$.

\subsection{Safety Specifications}

Safety specifications seem to be typically the simplest of the three specification types handled by \textsf{IMPaCT}. However, the interval iteration algorithm for safety specifications is not entirely  straightforward. In particular, for controller synthesis via finite MDPs, value iteration is used where the transition probability matrix $\hat{T}$ is multiplied by a vector of ones $V_1$ at each iteration backward over time, \emph{e.g.} $V_1' = \hat{T}V_1$.  In contrast, interval iteration in IMDP requires a static affine vector (\emph{i.e.,} either $\hat{A}$ or $\hat{R}$), otherwise
$V_0'$ will always remain $\textbf{0}_n$ as $V_0' = \hat{T}V_0$ with $V_0 = \textbf{0}_n$, and $V_1' = \hat{T}V_1$ with $V_0 = \textbf{1}_n$ at the start of the Bellman equation. Hence, the algorithm cannot converge, except when $V_1$ converges to $\textbf{0}_n$, indicating zero safety guarantee.

To resolve this challenge, given that safety and reachability are complement of each other, we fulfill our safety specification by solving a reachability problem over the complement of the safe set.
Logically speaking, $\square \mathcal{S}$ is equivalent to $\neg\lozenge \neg \mathcal{S}$, where $\neg$ refers to the LTL term ``not''. 
Therefore, the dynamic program to be solved is~\eqref{eq:linear-program}, where $\delta_1 = 0$, $\delta_2 = 1$, and $\delta_3 = V_1$, and the set $\mathcal{T}=\emptyset$. Additionally, we minimize with respect to the input $\hat{u}$ and maximize with respect to the disturbance $\hat{w}$. This approach stems from the intuition behind solving a reachability problem toward an \emph{undesired} area. Consequently, we aim to select the input least likely to reach the avoid region, which equates to choosing the input most likely to remain within the safe region. Similarly, the adversary selects the disturbance most likely to reach the avoid region. When the algorithm converges, the complement is taken to give the resulting $\mathds {P}_{\psi_{\min}}$ and $\mathds {P}_{\psi_{\max}}$. It is notable that for safety specifications over an \emph{infinite time horizon}, it is not uncommon for the solution to converge to the trivial result where $\mathds{P}_{\psi_{\min}} = 0$ and $\mathds{P}_{\psi_{\max}} = 1$, or for both bounds to converge to 0.

\begin{algorithm}[h!]
	\caption{Serial (Pessimistic) Safety Controller Synthesis}
	\label{alg:interval-safety-serial}
	\SetAlgoLined
	\KwData{IMDP $\hat{\Sigma}$, stopping condition $\varepsilon$}
	$V_0 := \textbf{0}_n$, $V_1 := \textbf{1}_n$, $\pi := \textbf{0}_n$\;
	\While{$\lVert V_1 - V_0\rVert{_{\infty}} > \varepsilon$}{
		\For{$\hat{x}_k\in\mathcal{S},k=\{0,\ldots,\vert\mathcal{S}\vert\}$}{
			\textbf{Solve max LP}~\eqref{eq:linear-program} for $\hat{A}$ and $\hat{T}$\;
			$\pi(\hat{x}_k) :=  \argmin\limits_{\hat{u}\in\hat{U}}\max\limits_{\hat{w}\in\hat{W}}{\max\limits_{\hat{A},\hat{T}}}\{\hat{A}(\hat{x}_k,\cdot,\cdot) + \sum\limits_{\forall\hat{x}'\in\mathcal{S}}\hat{T}(\cdot\vert \hat{x}_k,\cdot,\cdot)V_0(\hat{x}')\}$\;
			$V_0'(\hat{x}_k) :=  \min\limits_{\hat{u}\in\hat{U}}\max\limits_{\hat{w}\in\hat{W}}{\max\limits_{\hat{A},\hat{T}}}\{\hat{A}(\hat{x}_k,\cdot,\cdot) + \sum\limits_{\forall\hat{x}'\in\mathcal{S}}\hat{T}(\cdot\vert \hat{x}_k,\cdot,\cdot)V_0(\hat{x}')\}$\;
			$V_1'(\hat{x}_k) :=  \min\limits_{\hat{u}\in\hat{U}}\max\limits_{\hat{w}\in\hat{W}}{\max\limits_{\hat{A},\hat{T}}}\{\hat{A}(\hat{x}_k,\cdot,\cdot) + \sum\limits_{\forall\hat{x}'\in\mathcal{S}}\hat{T}(\cdot\vert \hat{x}_k,\cdot,\cdot)V_1(\hat{x}')\}$\;
		}
		$V_0 := V_0',~V_1 := V_1'$\;
	}
	$\mathds {P}_{\psi_{\min}} = \textbf{1}_n - V_0$\;
	$V_0 := \textbf{0}_n$, $V_1 := \textbf{1}_n$\;
	
	\While{$\lVert V_1 - V_0\rVert{_{\infty}} > \varepsilon$}{
		\For{$\hat{x}_k\in\mathcal{S},k=\{0,\ldots,\vert\mathcal{S}\vert\}$}{
			\textbf{Solve min LP}~\eqref{eq:linear-program} for $\hat{A}$ and $\hat{T}$\;
			fix $\hat{u}$ from $\pi(\hat{x}_k)$\;
			$V_0'(\hat{x}_k) :=  \max\limits_{\hat{w}\in\hat{W}}{\min\limits_{\hat{A},\hat{T}}}\{\hat{A}(\hat{x}_k,\hat{u},\cdot) + \sum\limits_{\forall\hat{x}'\in\mathcal{S}}\hat{T}(\cdot\vert \hat{x}_k,\hat{u},\cdot)V_0(\hat{x}')\}$\;
			$V_1'(\hat{x}_k) :=  \max\limits_{\hat{w}\in\hat{W}}{\min\limits_{\hat{A},\hat{T}}}\{\hat{A}(\hat{x}_k,\hat{u},\cdot) + \sum\limits_{\forall\hat{x}'\in\mathcal{S}}\hat{T}(\cdot\vert \hat{x}_k,\hat{u},\cdot)V_1(\hat{x}')\}$\;
		}
		$V_0 := V_0',~V_1 := V_1'$\;
	}
	$\mathds {P}_{\psi_{\max}} = \textbf{1}_n - V_1$\;
	\KwResult{controller $\mathcal{C} = (\mathcal{S},\pi,\mathds {P}_{\psi_{\min}},\mathds {P}_{\psi_{\max}})$}
\end{algorithm}

In Algorithm~\ref{alg:interval-safety-serial}, {controller synthesis for a pessimistic safety specification in serial is displayed.}
Steps $6-7$, and Steps $17-18$ show the interval iteration algorithm where the first loops find the optimal \emph{pessimistic} control policy $\pi$ using the lower bound and the second loops fix the control policy to find the IMDP upper bound. Notice that when calculating $\mathds {P}_{\psi_{\min}}$, we solve the max LP, and for $\mathds {P}_{\psi_{\max}}$ we solve the min LP.
Step $12$ and Step $22$ show the complement is taken, giving us the safety probability bounds.

\begin{algorithm}[h!]
	\caption{Serial (Pessimistic) Reach-Avoid Controller Synthesis}
	\label{alg:interval-reach-avoid-serial}
	\SetAlgoLined
	\KwData{IMDP $\hat{\Sigma}$, stopping condition $\varepsilon$}
	$V_0 := \textbf{0}_n$, $V_1 := \textbf{1}_n$, $\pi := \textbf{0}_n$\;
	\While{$\lVert V_1 - V_0\rVert{_{\infty}} > \varepsilon$}{
		\For{$\hat{x}_k\in\mathcal{S},k=\{0,\ldots,\vert\mathcal{S}\vert\}$}{
			\textbf{Solve min LP}~\eqref{eq:linear-program} for $\hat{R}$ and $\hat{T}$\;
			$\pi(\hat{x}_k) :=   \argmax\limits_{\hat{u}\in\hat{U}}\min\limits_{\hat{w}\in\hat{W}}{\min\limits_{\hat{R},\hat{T}}}\{\hat{R}(\hat{x}_k,\cdot,\cdot) + \sum\limits_{\forall\hat{x}'\in\hat{X}}\hat{T}(\cdot\vert \hat{x}_k,\cdot,\cdot)V_0(\hat{x}')\}$\;
			$V_0'(\hat{x}_k) :=  \max\limits_{\hat{u}\in\hat{U}}\min\limits_{\hat{w}\in\hat{W}}{\min\limits_{\hat{R},\hat{T}}}\{\hat{R}(\hat{x}_k,\cdot,\cdot) + \sum\limits_{\forall\hat{x}'\in\hat{X}}\hat{T}(\cdot\vert \hat{x}_k,\cdot,\cdot)V_0(\hat{x}')\}$\;
			$V_1'(\hat{x}_k) :=  \max\limits_{\hat{u}\in\hat{U}}\min\limits_{\hat{w}\in\hat{W}}{\min\limits_{\hat{R},\hat{T}}}\{\hat{R}(\hat{x}_k,\cdot,\cdot) + \sum\limits_{\forall\hat{x}'\in\hat{X}}\hat{T}(\cdot\vert \hat{x}_k,\cdot,\cdot)V_1(\hat{x}')\}$\;
		}
		$V_0 := V_0',~V_1 := V_1'$\;
	}
	$\mathds {P}_{\psi_{\min}} = V_0$\;
	$V_0 := \textbf{0}_n$, $V_1 := \textbf{1}_n$\;
	\While{$\lVert V_1 - V_0\rVert{_{\infty}} > \varepsilon$}{
		\For{$\hat{x}_k\in\mathcal{S},k=\{0,\ldots,\vert\mathcal{S}\vert\}$}{
			\textbf{Solve max LP}~\eqref{eq:linear-program} for $\hat{R}$ and $\hat{T}$\;
			fix $\hat{u}$ from $\pi(\hat{x}_k)$\;
			$V_0'(\hat{x}_k) :=  \min\limits_{\hat{w}\in\hat{W}}{\max\limits_{\hat{R},\hat{T}}}\{\hat{R}(\hat{x}_k,\hat{u},\cdot) + \sum\limits_{\forall\hat{x}'\in\hat{X}}\hat{T}(\cdot\vert \hat{x}_k,\hat{u},\cdot)V_0(\hat{x}')\}$\;
			$V_1'(\hat{x}_k) :=  \min\limits_{\hat{w}\in\hat{W}}{\max\limits_{\hat{R},\hat{T}}}\{\hat{R}(\hat{x}_k,\hat{u},\cdot) + \sum\limits_{\forall\hat{x}'\in\hat{X}}\hat{T}(\cdot\vert \hat{x}_k,\hat{u},\cdot)V_1(\hat{x}')\}$\;
		}
		$V_0 := V_0',~V_1 := V_1'$\;
	}
	$\mathds {P}_{\psi_{\max}} = V_1$\;
	\KwResult{controller $\mathcal{C} = (\mathcal{S},\pi,\mathds {P}_{\psi_{\min}},\mathds {P}_{\psi_{\max}})$}
\end{algorithm}

\subsection{Reachability and Reach-while-Avoid Specifications}

Reachability specifications typically introduce additional complexities compared to safety specifications, primarily due to the set of target states that the control policy must guide the system toward. Specifically, states that are not in the target (or avoid) region are considered safe, but the system should make progress towards the target set $\mathcal{T}$. Interval iteration for reachability is relatively straightforward. In our setting, $\hat{R}$ represents the transition probability of a state to the target region $\mathcal{T}$.

Reach-while-avoid specifications consider the case where $\hat{a}\in\mathcal{A}$ in the state space should be avoided. Then the dynamic program to be solved is~\eqref{eq:linear-program}, where $\delta_1 = 1$, $\delta_2 = 0$, and $\delta_3 = V_1$. When the algorithm converges, the resulting solution is $\mathds {P}_{\psi_{\min}}$ and $\mathds {P}_{\psi_{\max}}$ for each bound. It is worth noting that the synthesis of a reachability specification is equivalent to a reach-while-avoid specification where the avoid region $\mathcal{A} = \emptyset$.

A serial pessimistic reach-while-avoid controller approach is outlined in Algorithm~\ref{alg:interval-reach-avoid-serial}, in which the interval iteration occurs in Steps $6-7$ and Steps $17-18$. The first \textbf{while} loop iterates to synthesize a control policy $\pi$, and the second \textbf{while} loop uses $\pi$ by fixing the input $\hat{u}$ for each state $\hat{x}_k$. Steps $3$ and $14$ show the reach-avoid specification where the avoid set is removed from the state space leaving only $\hat{x}\in\mathcal{S}$. As earlier mentioned, the algorithm remains equivalent for the simpler reachability specification when $\mathcal{A} = \emptyset$. Steps $11$ and $22$ store the minimal and maximal probabilities of satisfaction corresponding to the policy $\pi$. Compared with the safety algorithm, no complement is required when calculating those probabilities.

\begin{remark}
	\textsf{IMPaCT} can also support finite-horizon specifications within IMDP construction by replacing the \textbf{while} loop in Algorithms~\ref{alg:interval-safety-serial} and~\ref{alg:interval-reach-avoid-serial} with a \textbf{for} loop over a finite number of time intervals. If this scenario applies, we still calculate each bound separately, fixing the policy $\pi$ for the second bound. In this case, interval iteration can be reduced to value iteration, enabling the solution of finite-horizon problems, albeit with the trade-off of sacrificing convergence guarantees~\cite{haddad2018interval}.
\end{remark}

\begin{remark}\label{rem:absorbing}
	In certain scenarios, such as the presence of absorbing states beyond those already outlined in the target and/or avoid regions, the two bounds of the interval iteration algorithm might not converge. In this case, calculating the number of time intervals $k$ for which each independent bound of the interval iteration algorithm converges to itself (with a small enough threshold)
	can be beneficial. The larger $k$ value can then be employed by the value iteration algorithm as its time horizon to hopefully ascertain a converged solution, ensuring the correctness of convergence. \textsf{IMPaCT} offers detailed guidance on the required steps to follow in such cases, including the example \verb$ex_load_safe$.
\end{remark}

\section{Parallel Controller Synthesis with Interval Iteration}
\label{sec:IMDPSynthesisParallel}

As offered in our approach, we can improve the computational efficiency of solving synthesis problems to acquire the resulting controller $\mathcal{C} = (\mathcal{S},\pi,\mathds {P}_{\psi_{\min}},\mathds {P}_{\psi_{\max}})$ by using parallel processing. The synthesis includes solving several dynamic programs and two interval iteration algorithms. We use the GNU linear programming kit~\cite{makhorin2008glpk} for solving the linear programs.

\begin{algorithm}[h!]
	\caption{\emph{Parallel} (Pessimistic) Safety Controller Synthesis}
	\label{alg:interval-safety}
	\SetAlgoLined
	\KwData{IMDP $\hat{\Sigma}$, stopping condition $\varepsilon$}
	$V_0 := \textbf{0}_n$, $V_1 := \textbf{1}_n$, $\pi := \textbf{0}_n$\;   
	\While{$\lVert V_1 - V_0\rVert{_{\infty}} > \varepsilon$}{
		\For{$x\in\mathcal{S}$ \textbf{in parallel}}{
			\textbf{Solve max LP}~\eqref{eq:linear-program} for $\hat{A}$ and $\hat{T}$\;
			$\pi(\hat{x}_k) :=  \argmin\limits_{\hat{u}\in\hat{U}}\max\limits_{\hat{w}\in\hat{W}}{\max\limits_{\hat{A},\hat{T}}}\{\hat{A}(\hat{x}_k,\cdot,\cdot) + \sum\limits_{\forall\hat{x}'\in\hat{X}}\hat{T}(\cdot\vert \hat{x}_k,\cdot,\cdot)V_0(\hat{x}')\}$\;
			$V_0'(\hat{x}_k) :=  \min\limits_{\hat{u}\in\hat{U}}\max\limits_{\hat{w}\in\hat{W}}{\max\limits_{\hat{A},\hat{T}}}\{\hat{A}(\hat{x}_k,\cdot,\cdot) + \sum\limits_{\forall\hat{x}'\in\hat{X}}\hat{T}(\cdot\vert \hat{x}_k,\cdot,\cdot)V_0(\hat{x}')\}$\;
			$V_1'(\hat{x}_k) :=  \min\limits_{\hat{u}\in\hat{U}}\max\limits_{\hat{w}\in\hat{W}}{\max\limits_{\hat{A},\hat{T}}}\{\hat{A}(\hat{x}_k,\cdot,\cdot) + \sum\limits_{\forall\hat{x}'\in\hat{X}}\hat{T}(\cdot\vert \hat{x}_k,\cdot,\cdot)V_1(\hat{x}')\}$\;
		}
		$V_0 := V_0',~V_1 := V_1'$\;
	}
	$\mathds {P}_{\psi_{\min}} = \textbf{1}_n - V_0$\;
	$V_0 := \textbf{0}_n$, $V_1 := \textbf{1}_n$\;
	
	\While{$\lVert V_1 - V_0\rVert{_{\infty}} > \varepsilon$}{
		\For{$x\in\mathcal{S}$ \textbf{in parallel}}{
			\textbf{Solve min LP}~\eqref{eq:linear-program} for $\hat{A}$ and $\hat{T}$\;
			fix $\hat{u}$ from $\pi(\hat{x}_k)$\;
			$V_0'(\hat{x}_k) :=  \max\limits_{\hat{w}\in\hat{W}}{\min\limits_{\hat{A},\hat{T}}}\{\hat{A}(\hat{x}_k,\hat{u},\cdot) + \sum\limits_{\forall\hat{x}'\in\hat{X}}\hat{T}(\cdot\vert \hat{x}_k,\hat{u},\cdot)V_0(\hat{x}')\}$\;
			$V_1'(\hat{x}_k) :=  \max\limits_{\hat{w}\in\hat{W}}{\min\limits_{\hat{A},\hat{T}}}\{\hat{A}(\hat{x}_k,\hat{u},\cdot) + \sum\limits_{\forall\hat{x}'\in\hat{X}}\hat{T}(\cdot\vert \hat{x}_k,\hat{u},\cdot)V_1(\hat{x}')\}$\;
		}
		$V_0 := V_0',~V_1 := V_1'$\;
	}
	$\mathds {P}_{\psi_{\max}} = \textbf{1}_n - V_1$\;
	\KwResult{controller $\mathcal{C} = (\mathcal{S},\pi,\mathds {P}_{\psi_{\min}},\mathds {P}_{\psi_{\max}})$}
\end{algorithm}

\begin{algorithm}[h!]
	\caption{\emph{Parallel} (Pessimistic) Reach-Avoid Controller Synthesis}
	\label{alg:interval-reach-avoid}
	\SetAlgoLined
	\KwData{IMDP $\hat{\Sigma}$, stopping condition $\varepsilon$}
	$V_0 := \textbf{0}_n$, $V_1 := \textbf{1}_n$, $\pi := \textbf{0}_n$\;
	\While{$\lVert V_1 - V_0\rVert{_{\infty}} > \varepsilon$}{
		\For{$x\in\mathcal{S}$ \textbf{in parallel}}{
			\textbf{Solve min LP}~\eqref{eq:linear-program} for $\hat{R}$ and $\hat{T}$\;
			$\pi(\hat{x}_k) :=   \argmax\limits_{\hat{u}\in\hat{U}}\min\limits_{\hat{w}\in\hat{W}}{\min\limits_{\hat{R},\hat{T}}}\{\hat{R}(\hat{x}_k,\cdot,\cdot) + \sum\limits_{\forall\hat{x}'\in\hat{X}}\hat{T}(\cdot\vert \hat{x}_k,\cdot,\cdot)V_0(\hat{x}')\}$\;
			$V_0'(\hat{x}_k) :=  \max\limits_{\hat{u}\in\hat{U}}\min\limits_{\hat{w}\in\hat{W}}{\min\limits_{\hat{R},\hat{T}}}\{\hat{R}(\hat{x}_k,\cdot,\cdot) + \sum\limits_{\forall\hat{x}'\in\hat{X}}\hat{T}(\cdot\vert \hat{x}_k,\cdot,\cdot)V_0(\hat{x}')\}$\;
			$V_1'(\hat{x}_k) :=  \max\limits_{\hat{u}\in\hat{U}}\min\limits_{\hat{w}\in\hat{W}}{\min\limits_{\hat{R},\hat{T}}}\{\hat{R}(\hat{x}_k,\cdot,\cdot) + \sum\limits_{\forall\hat{x}'\in\hat{X}}\hat{T}(\cdot\vert \hat{x}_k,\cdot,\cdot)V_1(\hat{x}')\}$\;
		}
		$V_0 := V_0',~V_1 := V_1'$\;
	}
	$\mathds {P}_{\psi_{\min}} = V_0$\;
	$V_0 := \textbf{0}_n$, $V_1 := \textbf{1}_n$\;
	\While{$\lVert V_1 - V_0\rVert{_{\infty}} > \varepsilon$}{
		\For{$x\in\mathcal{S}$ \textbf{in parallel}}{
			\textbf{Solve max LP}~\eqref{eq:linear-program} for $\hat{R}$ and $\hat{T}$\;
			fix $\hat{u}$ from $\pi(\hat{x}_k)$\;
			$V_0'(\hat{x}_k) :=  \min\limits_{\hat{w}\in\hat{W}}{\max\limits_{\hat{R},\hat{T}}}\{\hat{R}(\hat{x}_k,\hat{u},\cdot) + \sum\limits_{\forall\hat{x}'\in\hat{X}}\hat{T}(\cdot\vert \hat{x}_k,\hat{u},\cdot)V_0(\hat{x}')\}$\;
			$V_1'(\hat{x}_k) :=  \min\limits_{\hat{w}\in\hat{W}}{\max\limits_{\hat{R},\hat{T}}}\{\hat{R}(\hat{x}_k,\hat{u},\cdot) + \sum\limits_{\forall\hat{x}'\in\hat{X}}\hat{T}(\cdot\vert \hat{x}_k,\hat{u},\cdot)V_1(\hat{x}')\}$\;
		}
		$V_0 := V_0',~V_1 := V_1'$\;
	}
	$\mathds {P}_{\psi_{\max}} = V_1$\;
	\KwResult{controller $\mathcal{C} = (\mathcal{S},\pi,\mathds {P}_{\psi_{\min}},\mathds {P}_{\psi_{\max}})$}
\end{algorithm}

\subsection{Implementations}
\label{sec:CPUsynth}
The performance of the safety, reachability, and reach-avoid specifications can be enhanced by running in parallel all of the nested \textbf{for} loops from Algorithms~\ref{alg:interval-safety-serial} and~\ref{alg:interval-reach-avoid-serial}. These can be seen in the updated Algorithm~\ref{alg:interval-safety} in Step $3$, and Step $14$ for safety specifications. Once more, it is crucial to highlight that we aim to synthesize the minimal $\hat{u}$ as we seek the input that minimizes the likelihood of reaching the region to avoid (\emph{i.e.,} complement of the safe set).
Algorithm~\ref{alg:interval-reach-avoid} presents an updated parallel implementation of Algorithm~\ref{alg:interval-reach-avoid-serial} for reach-while-avoid specifications. In particular, Step 3 and Step 14 replace the previous \textbf{for} loops in parallel. We reiterate that when $\mathcal{A}=\emptyset$, the reach-avoid algorithm is reduced to a simpler reachability algorithm.

\textsf{IMPaCT} automatically detects whether to use reachability or reach-avoid specification algorithms based on whether an avoid region has been set by the user. Additionally a Boolean value, \verb$IMDP_lower$, is set as \verb$true$ for a pessimistic policy, or \verb$false$ for an optimistic policy. For finite-horizon controllers, an additional parameter is needed for the time horizon.

\begin{lstlisting}[language=C++,caption=Synthesis Algorithms.]
void setStoppingCondition(double eps);
void infiniteHorizonReachController(bool IMDP_lower);
void infiniteHorizonSafeController(bool IMDP_lower);
void finiteHorizonReachController(bool IMDP_lower, size_t timeHorizon);
void finiteHorizonSafeController(bool IMDP_lower, size_t timeHorizon);
\end{lstlisting}

\subsection{Controller Synthesis over GPU} \textsf{IMPaCT} enables the utilization of GPUs for the verification and synthesis purposes via the constructed IMCs/IMDPs. The implementation leverages the \emph{sorting method} from~\cite[Lemma $7$]{sen2006model} and provides the same guarantees as the LP solving approach described in Section~\ref{sec:CPUsynth}. The primary computational advantage lies in the absence of external library functions such as GLPK. Consequently, lines $4$ and $15$ in Algorithm~\ref{alg:interval-safety} and Algorithm~\ref{alg:interval-reach-avoid} can be substituted with a sorting method to compute the feasible distribution. Since the algorithms can be implemented manually without relying on external functions, the GPU can be leveraged for these computations. Intuitively, the sorting method solves the LP from~\eqref{eq:linear-program} by initially defining the feasible distribution as the minimum transition probabilities, employing $\hat{T}_{\min}(\hat{x}'\vert\hat{x},\hat{u},\hat{w})$, $\hat{A}_{\min}(\hat{x},\hat{u},\hat{w})$ and $\hat{R}_{\min}(\hat{x},\hat{u},\hat{w})$. If the sum of these probabilities is less than $1$, then each value of the feasible distribution is sequentially updated until the sum equals $1$. The probability of a single value can be increased until it reaches $\hat{T}_{\max}(\hat{x}'\vert\hat{x},\hat{u},\hat{w})$, $\hat{A}_{\min}(\hat{x},\hat{u},\hat{w})$ or $\hat{R}_{\min}(\hat{x},\hat{u},\hat{w})$, respectively. Subsequently, the next value in the ordering will be incremented until the total sum equals $1$. The ordering is determined by the values $\delta_1$, $\delta_2$, and $\delta_3(\hat{x}')$, with the maximal LP employing a \emph{descending} order and the minimal LP employing an \emph{ascending} order.

\begin{lstlisting}[language=C++,caption=Sorted Synthesis Algorithms.]
void infiniteHorizonReachControllerSorted(bool IMDP_lower);
void infiniteHorizonSafeControllerSorted(bool IMDP_lower);
void finiteHorizonReachControllerSorted(bool IMDP_lower, size_t timeHorizon);
void finiteHorizonSafeControllerSorted(bool IMDP_lower, size_t timeHorizon);
\end{lstlisting}

With the sorted approach, synthesis can be readily computed by appending \verb$Sorted$ to the desired synthesis function. As demonstrated in Table~\ref{tab:time} and Table~\ref{tab:GPU}, the sorted approach yields significant performance improvements for both CPU and GPU computations across various example classes.

\begin{remark}
In the current implementation, the sorting process itself is not parallelized and utilizes \verb$std::sort$, while subsequent computations are parallelized. Implementing a parallel sorting algorithm manually could be regarded as a potential future extension to \textsf{IMPaCT}. It is worth noting that for large systems, the sorting method still offers substantial benefits over the LP solver.
\end{remark}

\begin{remark}
    The functions for synthesis along with the functions for saving and loading files, described in the next section, are located in the source file \verb$GPU_synthesis.cpp$, which is included by the main source file \verb$IMDP.cpp$. This separation has been intentionally implemented to prevent compilation errors when employing the GPU approach. For GPU utilization, the configuration file should be set to load all required matrices and vectors before invoking the sorted approach. The Makefile needs to be modified to ensure that the \verb$--acpp-targets$ flag specifies the GPU architecture, such as \verb$cuda:sm_80$. Furthermore, in the Makefile, the reference to the file \verb$IMDP.cpp$ should be substituted with \verb$GPU_synthesis.cpp$. For an illustration, we refer to the example \verb$ex_GPU$.
\end{remark}

\subsection{Verification Analysis via Interval Markov Chains (IMCs)}

In striving for maximum user-friendliness, the tool also automatically distinguishes between verification tasks using an IMC and controller synthesis via an IMDP. In particular, for verification analysis, users only need to specify the absence of input parameters, denoted by \verb$dim_u$$~=0$. Accordingly, functions like \verb$infiniteHorizonReachController$, etc., will identify the lack  of input and generate a lookup table represented as $\mathcal{C}:= (\mathcal{S},\mathds {P}_{\psi_{\min}},\mathds {P}_{\psi_{\max}})$, although it will still be saved as \verb$controller.h5$. In fact, the verification's lookup table provides both the minimum and maximum probabilities of fulfilling a desired property corresponding to each partition of the finite system.

\section{Loading and Saving Files}
\label{sec:load&save}
We use HDF5~\cite{hdf5} as the data format for saving and loading into \textsf{IMPaCT}. In particular, HDF5 is a common widely supported format for large, heterogeneous, and complex data sets. This data structure is self-descriptive, eliminating the need for extra metadata to interpret the files. Additionally, it supports ``data slicing", enabling extraction of specific segments from a dataset without the necessity of analyzing the entire set.

The primary advantage of HDF5 lies in its open format, which ensures native support across numerous programming languages and tools, such as MATLAB, Python, and R. This should facilitate simpler sharing of synthesized controllers without requiring end-users to install additional programs. In \textsf{IMPaCT}, the subsequent commands are employed to save data in HDF5 format:

\begin{lstlisting}[language=C++,caption=Save functions.]
void saveStateSpace();
void saveInputSpace();
void saveDisturbSpace();
void saveTargetSpace();
void saveAvoidSpace();
void saveMinTargetTransitionVector();
void saveMaxTargetTransitionVector();
void saveMinAvoidTransitionVector();
void saveMaxAvoidTransitionVector();
void saveMinTransitionMatrix();
void saveMaxTransitionMatrix();
void saveController();
\end{lstlisting}

Furthermore, users have the flexibility to build the abstraction using their preferred methods and load the required matrices into \textsf{IMPaCT} for verification or controller synthesis tasks.
For an illustration, we refer to the example \verb$ex_load_reach$.

\begin{lstlisting}[language=C++,caption=Load functions.]
void loadStateSpace(string filename);
void loadInputSpace(string filename);
void loadDisturbSpace(string filename);
void loadTargetSpace(string filename);
void loadAvoidSpace(string filename);
void loadMinTargetTransitionVectorx(string filename);
void loadMaxTargetTransitionVector(string filename);
void loadMinAvoidTransitionVector(string filename);
void loadMaxAvoidTransitionVector(string filename);
void loadMinTransitionMatrix(string filename);
void loadMaxTransitionMatrix(string filename);
void loadController(string filename);
\end{lstlisting}

\section{Benchmarking and Case Studies}
\label{sec:case_studies}

We illustrate \textsf{IMPaCT}'s applications by employing multiple well-known benchmark systems, encompassing safety, reachability, and reach-avoidance specifications across infinite horizons. Among these, we leverage several complex physical case studies, encompassing a 2D robot, a 3D autonomous vehicle, 3D and 5D room temperature control systems, as well as 4D and 7D building automation systems.
We also consider a 14D case study used for the purposes of showing the scalability of the tool. All the case studies are run on a high performance computer with 2 AMD EPYC 7702 64-Core along with 2TB RAM. In Table~\ref{tab:time}, we showcase the computational times for these case studies, all executed over an \emph{infinite time horizon}. In Table~\ref{tab:GPU}, we provide a comparison of various synthesis algorithms' performance on a standard desktop machine Linux, utilizing both CPU (12th Gen Intel Core i9-12900 $\times$ 24) and GPU (NVIDIA RTX A4000) configurations.

\subsection{2D Robot}
Consider a robot described by the following \emph{nonlinear} difference equations:
\begin{equation*}
\begin{bmatrix}
x_1(k+1)\\
x_2(k+1)
\end{bmatrix} = \begin{bmatrix}
x_1(k) + 10u_1(k)\cos(u_2(k)) + w(k) + \varsigma_1(k)\\
x_2(k) + 10u_2(k)\sin(u_2(k)) + w(k) + \varsigma_2(k)
\end{bmatrix}\!\!,
\end{equation*}
where $(x_1,x_2)\in X := [-10,10]^2$ is the spacial coordinate of the location of the robot, $(u_1,u_2)\in U := [-1,1]^2$ is the input, and $w\in W : = [-0.5,0.5]$ is the disturbance. The noise $(\varsigma_1,\varsigma_2)$ has the covariance matrix ${\mathsf{Cov}} := \mathsf{diag}(0.75,0.75)$. The discrete intervals are, respectively, $\eta_x = (0.5, 0.5)$, $\eta_u = (0.1, 0.1)$ and $\eta_w = 0.1$. The target set $\mathcal{T} := [5,7]^2$ and the avoid set $\mathcal{A} := [-2,2]^2$.

\begin{figure}
	\centering
	\begin{subfigure}{.49\textwidth}
		\centering
		\includegraphics[width=0.96\linewidth]{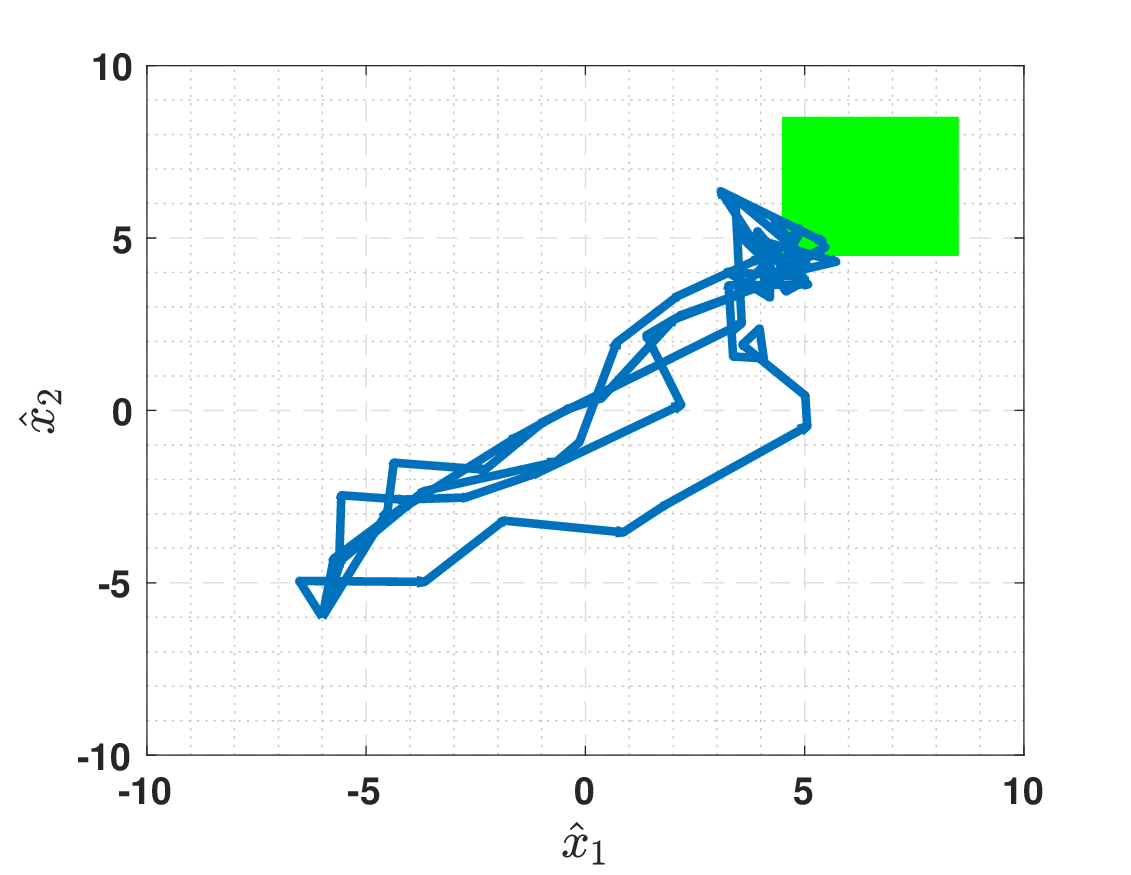}
		\caption{2D Robot - Reachability}
		\label{fig:2DRobot-sub1}
	\end{subfigure}\hspace{0.1cm}
	\begin{subfigure}{.49\linewidth}
		\centering
		\includegraphics[width=1.01\linewidth]{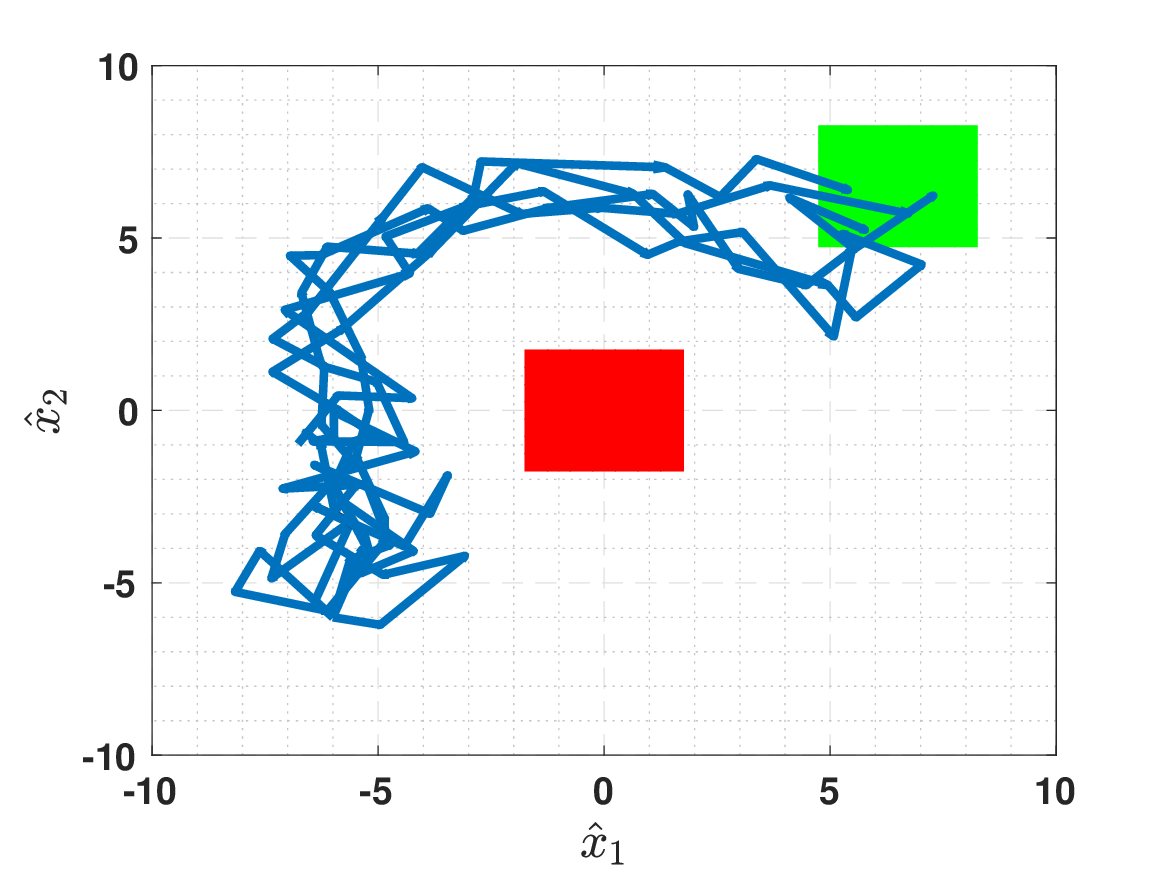}
		\caption{2D Robot - Reach-Avoid}
		\label{fig:2DRobot-sub2}
	\end{subfigure}
	\caption{2D Robot case study fulfilling reachability and reach-avoid properties with different noise realizations. The green and red boxes are target and avoid regions, respectively.}
	\label{fig:2DRobot}
\end{figure}

We consider four scenarios for this case study:
\begin{itemize}
	\item Reach-avoid specification with no disturbance, see Fig.~\ref{fig:2DRobot}(B), \emph{i.e.,} $\hat{w}(k) = 0$.
	\item Reach-avoid specification with disturbance.
	\item Reachability specification with no disturbance with updated $\eta_x = (1, 1)$ and $\eta_u = (0.2, 0.2)$, see Fig.~\ref{fig:2DRobot}(A).
	\item Reachability specification with disturbance with with updated $\eta_x = (1, 1)$, $\eta_u = (0.2, 0.2)$.
\end{itemize}
The latter two scenarios are designed for swift simulations on small personal computers.

\subsection{3D Autonomous Vehicle}
\begin{figure}
	\centering
	\includegraphics[width=0.5\linewidth]{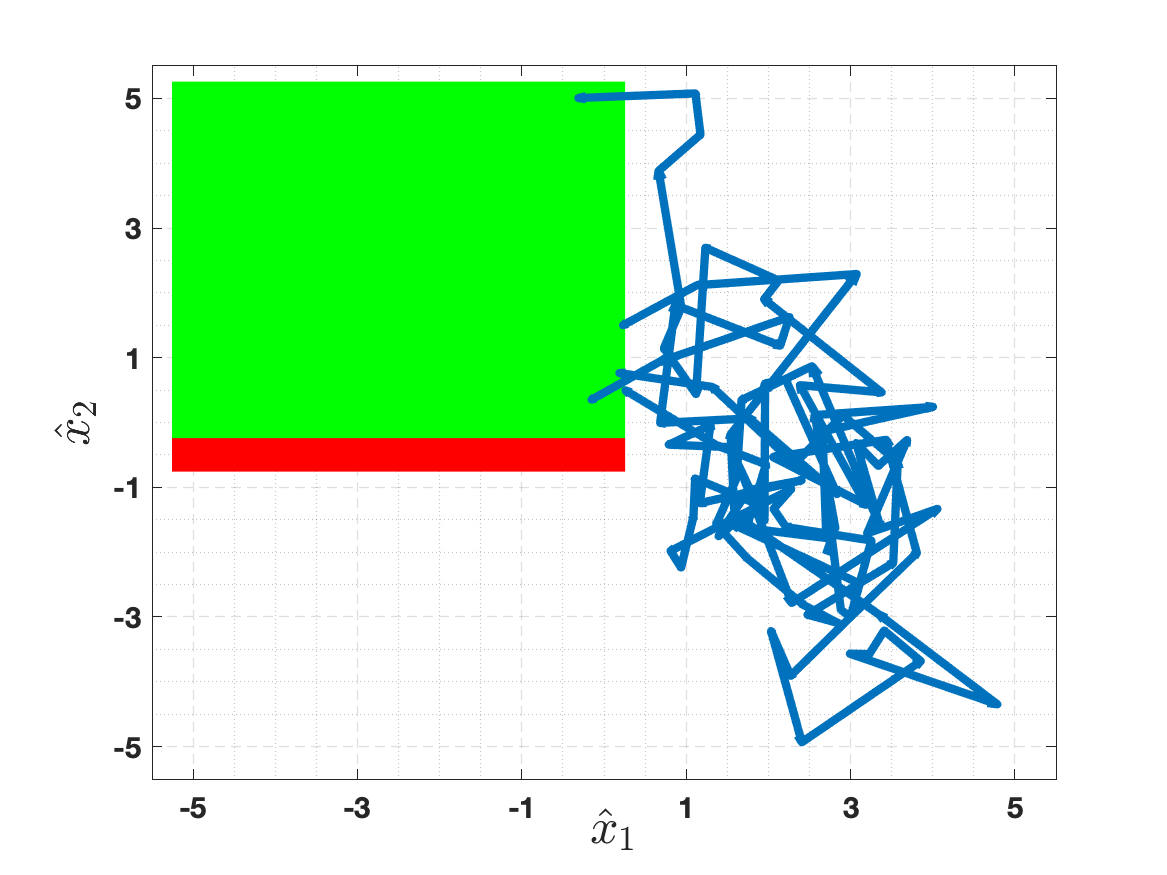}
	\caption{3D Autonomous Vehicle case study fulfilling reach-while-avoid property, with different noise realizations, starting from an initial condition $[3;-3;0.6]$. The green and red boxes are target and avoid regions, respectively.}
	\label{fig:3Dvehicle}
\end{figure}

Consider a 3-dimensional autonomous vehicle described by the following \emph{nonlinear} difference equations:
\begin{equation*}
\begin{bmatrix}
x_1(k+1)\\
x_2(k+1)\\
x_3(k+1)
\end{bmatrix} = 
\begin{bmatrix}
x_1(k) + (u_1(k)\cos(\alpha + x_3(k))\cos(\alpha)^{-1})T_s + {\varsigma_1(k)}\\
x_2(k) + (u_1(k)\sin(\alpha + x_3(k))\cos(\alpha)^{-1} )T_s + {\varsigma_2(k)}\\
x_3(k) + (u_1(k)\tan(u_2(k)) )T_s + {\varsigma_3(k)}
\end{bmatrix}\!\!,
\end{equation*}
where $\alpha = \arctan(\frac{\tan(u_2)}{2})$, and $T_s = 0.1$ is the sampling time. The inputs $(u_1,u_2)\in U := [-1,4]\times[-0.4,0.4]$ represent the wheel velocity and the steering angle, where $\eta_u = (1,0.2)$. The states $(x_1,x_2)$ are the spatial coordinates, and $x_3$ is the orientation, where $(x_1,x_2,x_3)\in X := [-5,5]^2\times [-3.4, 3.4]$ and $\eta_x = (0.5, 0.5, 0.4)$. The noise $(\varsigma_1,\varsigma_2,\varsigma_3)$ has covariance matrix ${\mathsf{Cov}} := \mathsf{diag}(\frac{2}{3},\frac{2}{3},\frac{2}{3})$. The target set $\mathcal{T}$ and avoid set $\mathcal{A}$ are described by the hyper-rectangles $[-5.75,0.25]\times[-0.25,5.75]\times[-3.45, 3.45]$ and $[-5.75,0.25]\times[-0.75,-0.25]\times[-3.45, 3.45]$, respectively. For this case study, we consider a reach-while-avoid specification with no disturbances, see Fig.~\ref{fig:3Dvehicle}. We also examine a larger version of the same case study, incorporating $\eta_u = (0.5,0.1)$ and $\eta_x = (0.5,0.5,0.2)$.

\subsection{3D and 5D Room Temperature Control Systems}

We also employ \textsf{IMPaCT} for the room temperature control systems taken from the ARCH competition~\cite{abate2020arch}, one with $3$ dimensions and another one with $5$ dimensions. The two case studies both consider two inputs described by $(u_1,u_2)\in U := [0,1]^2$ with $\eta_u = (0.2,0.2)$, and have a state space described by $(x_1, \ldots, x_n)\in X := [19,21]^{n}$, where $n$ is the state dimension. For the $3$D case, the dynamics are described as:
\begin{equation*}
\begin{bmatrix}
x_1(k+1)\\
x_2(k+1)\\
x_3(k+1)
\end{bmatrix} \!=\! 
\begin{bmatrix}
(a-\gamma u_1(k))x_1(k) + {\zeta}(x_2(k) + x_3(k)) + \gamma T_hu_1(k) + \beta T_e + \varsigma_1(k)\\
ax_2(k) + {\zeta}(x_1(k) + x_3(k)) + \beta T_e + \varsigma_2(k)\\
(a-\gamma u_2(k))x_3(k) + {\zeta}(x_1(k) + x_2(k)) + \gamma T_hu_2(k) + \beta T_e + \varsigma_3(k)
\end{bmatrix}\!\!,
\end{equation*}
where $\beta = 0.022, \gamma = 0.05, {\zeta}=0.2$, are the conduction factors, respectively, between the external environment and the current room, between the heater and the current room, and between neighbouring rooms. In addition, $T_h = 50^{\circ} C$ is the heater temperature, $T_e = -1^{\circ} C$ is the outside temperature, and $a = 1-2{\zeta} -\beta$. Furthermore, $\eta_x = (0.1,0.1,0.1)$ and the noise has covariance matrix ${\mathsf{Cov}} := \mathsf{diag}(0.02,0.02,0.02)$. 

\begin{figure}
	\centering
	\includegraphics[width=0.85\linewidth]{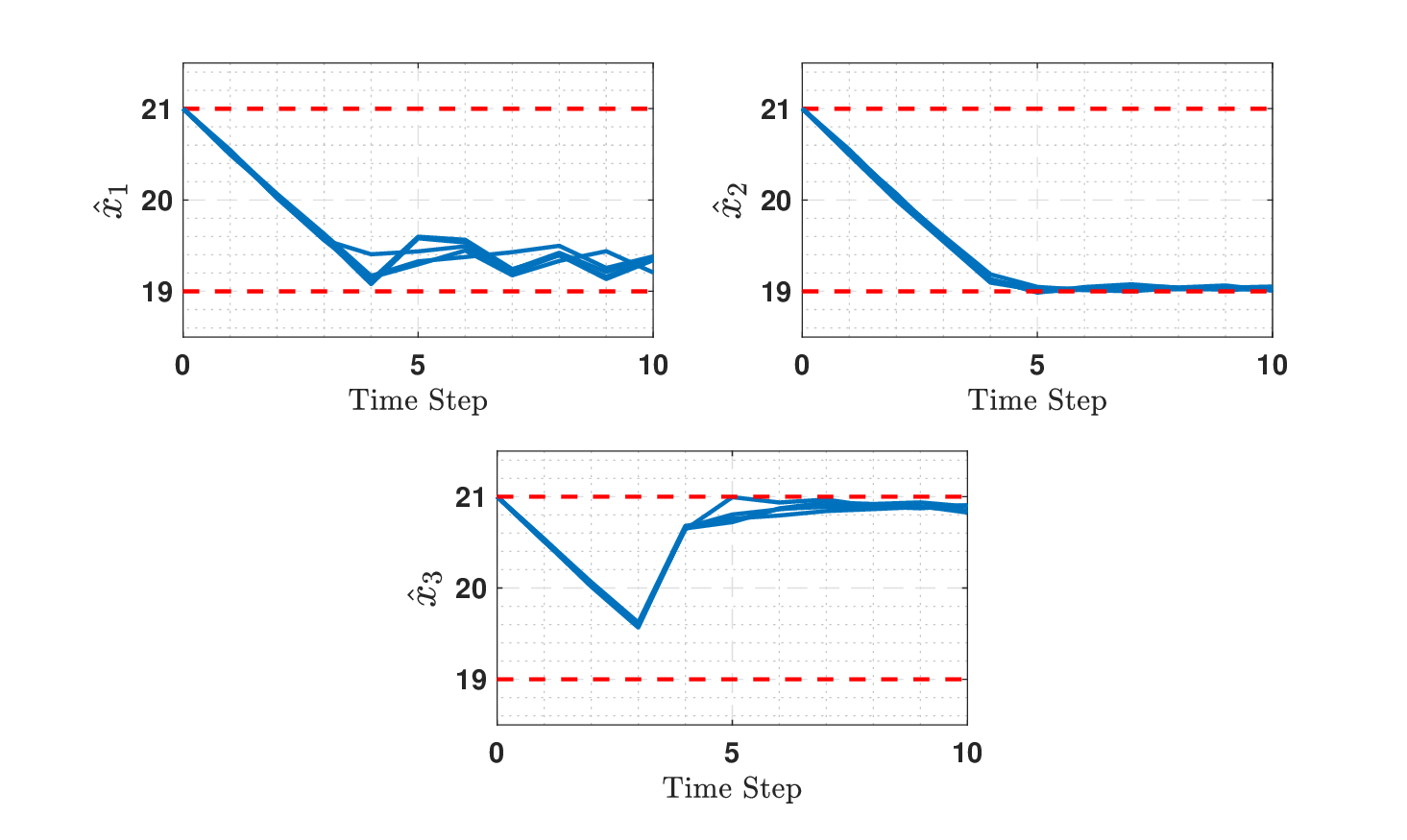}
	\caption{3D Room Temperature fulfilling safety properties over $10$ time steps, with different noise realizations, starting from an initial condition $[21;21;21]$.}
	\label{fig:3Droom}
\end{figure}

The dynamics for the $5$D case study are characterized as:
\begin{equation*}
\begin{bmatrix}
x_1(k+1)\\
x_2(k+1)\\
x_3(k+1) \\
x_4(k+1) \\
x_5(k+1) 
\end{bmatrix} \!=\!
\begin{bmatrix}
(a-\gamma u_1(k))x_1(k) + {\zeta}(x_2(k) + x_5(k)) + \gamma T_hu_1(k) + \beta T_e + \varsigma_1(k)\\
ax_2(k) + {\zeta}(x_1(k) + x_3(k)) + \beta T_e + \varsigma_2(k)\\
(a-\gamma u_2(k))x_3(k) + {\zeta}(x_4(k) + x_2(k)) + \gamma T_hu_2(k) + \beta T_e + \varsigma_3(k) \\
ax_4(k) + {\zeta}(x_3(k)+x_5(k))+\beta T_e +{\varsigma_4(k)}\\
ax_5(k) + {\zeta}(x_4(k)+x_1(k))+\beta T_e + {\varsigma_5(k)}
\end{bmatrix}\!\!,
\end{equation*} 
where ${\zeta}=0.3$ and $\eta_x = (0.4,0.4,0.4,0.4,0.4)$. The noise has covariance matrix ${\mathsf{Cov}} :=\mathsf{diag}(0.01,0.01,0.01,0.01,0.01)$.

Both case studies examine safety specifications where the safe region 
$\mathcal{S}=X$. Figure~\ref{fig:3Droom} showcases a simulation of the safety controller spanning $10$ time steps, while Table~\ref{tab:time} presents the synthesis times for this case study within an infinite time horizon.

\subsection{4D and 7D Building Automation Systems}

\begin{figure}
	\centering
	\includegraphics[width=0.88\linewidth]{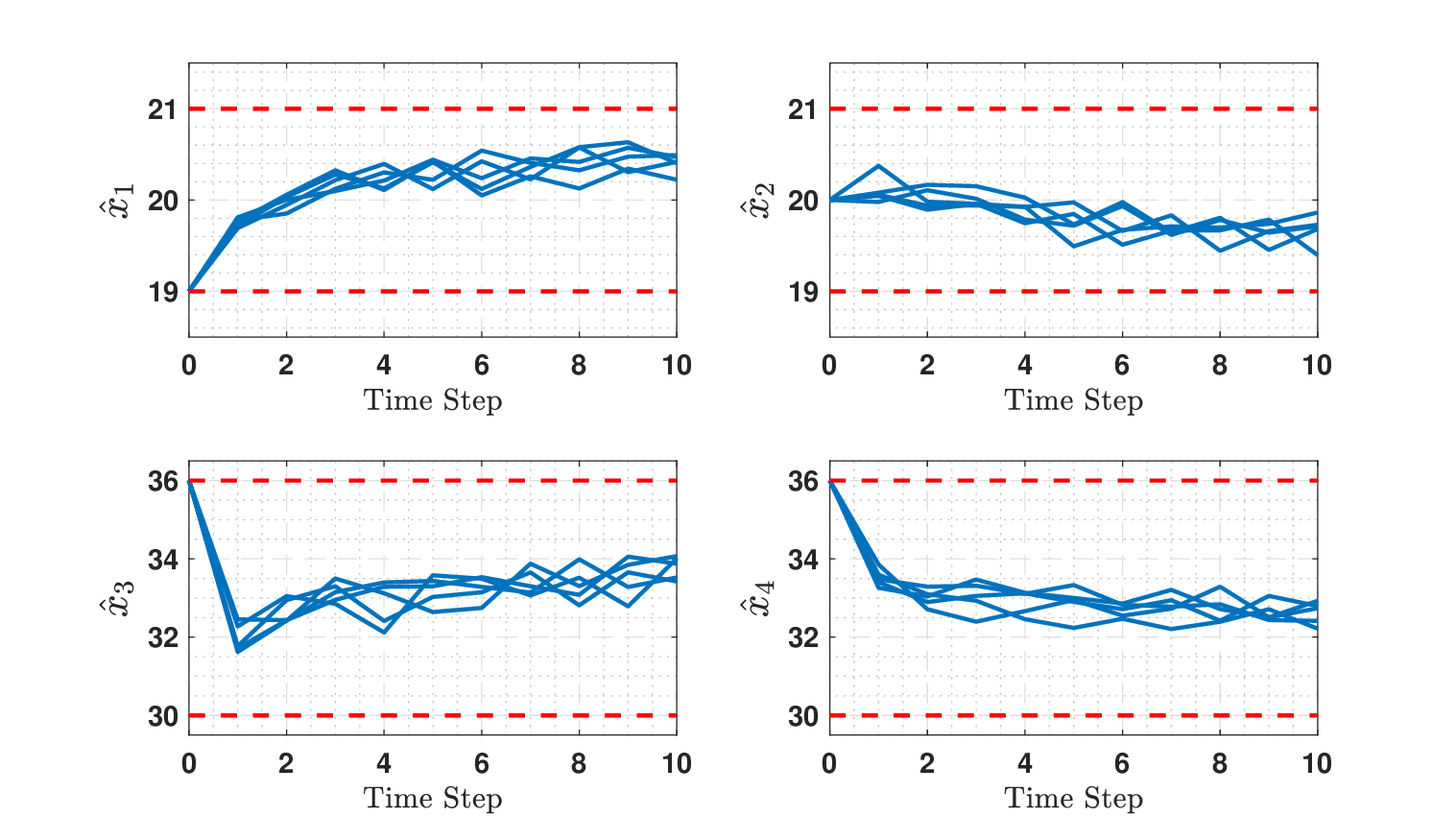}
	\caption{4D Building Automation System fulfilling safety properties within $10$ time steps, with $5$ different noise realizations, starting from an initial condition $[19; 20;36;36]$.}
	\label{fig:4DBAS}
\end{figure}

The following benchmarks are based on a smart building at the University of Oxford~\cite{cauchi2018benchmarks}. We consider two versions of this benchmark with different dimensions. The first one involves a safety synthesis problem rooted in a two-zone configuration, featuring stochastic dynamics:
\begin{align*}
x(k+1) = \begin{bmatrix}
0.6682 & 0 & 0.02632 & 0\\ 0 & 0.683 & 0 & 0.02096\\ 1.0005 & 0 &  -0.000499 & 0\\
0 & 0.8004 & 0 & 0.1996
\end{bmatrix} x(k) \nonumber+ \begin{bmatrix}
0.1320\\ 0.1402\\ 0\\ 0
\end{bmatrix} u(k) + \begin{bmatrix}
3.4378 \\ 2.9272\\ 13.0207\\ 10.4166
\end{bmatrix} + \varsigma(k),
\end{align*}
where the covariance matrix ${\mathsf{Cov}} := {\mathsf{diag}}(12.9199,12.9199,2.5826, 3.2279)$ and the off-diagonal elements are zero. Moreover, $x\in X := [19,21]^2\times[30,36]^2$ with $\eta_x := (0.5,0.5,1.0,1.0)$, and $u\in U := [17,20]$ with $\eta_u := (1.0,1.0)$, see Fig.~\ref{fig:4DBAS}.

The second version is a safety \emph{verification} problem (\emph{i.e.,} without input variables), in which the dynamics extend the first benchmark to consider a larger number of continuous variables:
\begin{align*}
x(k+1) = \begin{bmatrix}
0.968& 0& 0.004& 0& 0.004& 0& 0.004\\ 0& 0.968& 0& 0.003& 0& 0.003& 0.003\\ 0.011& 0& 0.949& 0& 0& 0& 0\\ 0& 0.010& 0& 0.952& 0& 0& 0\\ 0.011& 0& 0& 0& 0.949& 0& 0\\ 0& 0.010& 0& 0& 0& 0.952& 0\\ 0.011& 0.010& 0& 0& 0& 0& 0.979
\end{bmatrix} x(k) + \varsigma(k),
\end{align*}
where the noise has covariance ${\mathsf{Cov}}:={\mathsf{diag}}(51.3, 50.0, 21.8, 23.5, 25.2, 26.5, 91.7)$, with off-diagonals equal to zero. In addition, $x(k)\in X := [-0.5,0.5]^7$ with $\eta_x := (0.05,0.05,0.5,0.5,0.5,0.5,0.5)$.

It is noteworthy that, as indicated in Table~\ref{tab:time}, the construction of the IMDP abstraction for the $7$D building automation systems was accomplished within $24$ hours. However, the synthesis process extended beyond $24$ hours to complete. This prolonged duration is primarily attributed to the expansive size of the state space, comprising $107,163$ finite states, while each state functions as a decision variable within the dynamic program. Considering that the synthesis involves solving the dynamic program for each row, the number of decision variables in one iteration amounts to $107,163^2$, \emph{i.e.,} approximately $11.5$ billion decision variables.

\subsection{14D Case Study}

To demonstrate the scalability of \textsf{IMPaCT}, we consider a 14D case study, borrowed from the relevant literature~\cite{StocHy,AMYTISS}, with the following dynamics:
\begin{equation}
x(k+1) = 0.8x(k) + 0.2\varsigma(k), 
\end{equation}
where $x = [x_1;\dots;x_{14}], \varsigma = [\varsigma_1;\dots;\varsigma_{14}]$, with $X:=[-0.5,0.5]^{14}$. We consider this case study to solve a safety verification problem, as detailed in Table~\ref{tab:time}.

\subsection{Comparison with \textsf{StocHy}}

As previously highlighted in the introduction, \textsf{StocHy} utilizes the value iteration algorithm for IMDP construction, which \emph{lacks convergence guarantees} when dealing with infinite-horizon specifications. In contrast, \textsf{IMPaCT} employs the \emph{interval iteration} algorithm to guarantee convergence to an optimal controller. This crucial feature, in addition to offering parallelization, distinguishes \textsf{IMPaCT} from \textsf{StocHy}, which defaults to value iteration without general parallelization capabilities.

Beyond this guarantee, we aim to execute the $14$D case study using \textsf{StocHy} to showcase the efficiency of our tool. While \textsf{IMPaCT} completed the IMDP construction within $28.1$ minutes, as detailed in Table~\ref{tab:time}, \textsf{StocHy} failed to complete task within a $24$-hour timeframe.

\section{Conclusion}

{In this work, we developed the advanced software tool \textsf{IMPaCT}, which is the first tool to exclusively construct IMC/IMDP abstraction and perform verification and controller synthesis over \emph{infinite-horizon} properties while providing \emph{convergence guarantees}. Developed in C++ using AdaptiveCpp, an independent open-source implementation of SYCL, \textsf{IMPaCT} capitalizes on adaptive parallelism across diverse CPUs/GPUs of all hardware vendors, including Intel and NVIDIA. We have benchmarked \textsf{IMPaCT} across various physical case studies including a 2D robot, a 3D autonomous vehicle, a 5D room temperature control system, and a 7D building automation system, borrowed from the ARCH tool competition, with its scalability further highlighted through a 14D case study.
	
\section*{Acknowledgment}
The authors express their gratitude to Sadegh Soudjani for valuable discussions at the early stage of this project. Additionally, the authors would like to acknowledge Tobias Kaufmann in the Max Planck Institute for granting access to the high-performance computer utilized for simulations in Table~\ref{tab:time}.

\bibliographystyle{alpha}
\bibliography{sample-base}
\addtolength{\voffset}{-1.2cm}

\clearpage\thispagestyle{empty}

\begin{sidewaystable}
	\centering
	\vspace{12cm}\caption{\small
		Execution times and memory requirements of \textsf{IMPaCT} applied to a set of benchmarks. Computation times are in seconds and memory usages in MB, unless otherwise specified. Specifications: $\mathsf{S}$ for safety, $\mathsf{R}$ for reachability, and $\mathsf{R-A}$ for reach-while-avoid. BAS stands for Building Automation System. We signify the synthesis times using the GLPK Library with ``$^{a}$'' and the synthesis times based on the sorting method from~\cite[Lemma 7]{sen2006model} with ``$^{b}$''.
		Note that ``$^{*}$'' indicates possible absorbing states: in this case a finite horizon run is conducted with a convergent number of steps, where algorithm times are aggregated (see Remark~\ref{rem:absorbing}).
		In 4D BAS benchmark, using the GLPK library, controller synthesis for a finite horizon of $10$ steps is computed in $4.66$ seconds, while it takes over $6$ hours for the infinite horizon to converge.}
	{\footnotesize \begin{tabular}{|c|c|c|c|c|c|c|c|c|c|c|c|c|c|c|c|c|c|}
		Case Study & Spec & \multicolumn{4}{|c|}{} & \multicolumn{1}{|c}{$\hat{T}_{\min}$} & \multicolumn{1}{c}{$\hat{T}_{\max}$}& 
  \multicolumn{1}{c|}{}&
  \multicolumn{1}{|c}{$\hat{R}_{\min}$} &
  \multicolumn{1}{c}{$\hat{R}_{\max}$} &
  \multicolumn{1}{c|}{}&
		\multicolumn{1}{|c}{$\hat{A}_{\min}$} & \multicolumn{1}{c}{$\hat{A}_{\max}$} &
  \multicolumn{1}{c|}{}&
  \multicolumn{3}{|c|}{$\mathcal{C}$} \\
		\hline
		& &$\vert\hat{X}\vert$ &$\vert\hat{U}\vert$&$\vert\hat{W}\vert$&$\vert\hat{X}\times\hat{U}\times\hat{W}\vert$& time & time & mem & time  & time & mem & time & time & mem & {time$^{a}$} & {time$^{b}$} & mem\\
		\hline
		2D Robot & $\mathsf{R}$ & 441 & 121 & 0 & 53,361 & {0.60} & {1.58} & 174.8 & {0.09} & {0.294} & 4.5 & {0.016} & {0.015} & 4.5 & {7.34} & {8.53} & 0.02\\
		2D Robot & $\mathsf{R}$ & 441 & 121 & 11 & 586,971 & {5.9} & {15.6} & 1.9GB & {0.251} &  {1.10} & 6.58 & {0.04} & {0.04} & 6.58 & {65.7} & {330} & 0.02 \\
		2D Robot & $\mathsf{R-A}$ & 1,681 & 441 & 0 &741,321 & {20.7} & {59.8} & 8.8GB & {0.78}  & {2.51} & 61.4 & {1.86} & {1.88} & 61.4 & {1,549} & {1,047} & 0.08 \\
		2D Robot & $\mathsf{R-A}$ & 1,681 & 441 & 11 & 8,154,531 & {227} & {675} & 97.2GB & {7.46} & {22.0} & 274 & {21.2} & {22.6} & 274 & {5.66hr} & {8.46hr} & 0.08 \\
		3D Vehicle & $\mathsf{R-A}$ & 7,938 & 30 & 0 & 238,140 & {89.1} & {114} & 7.42GB & {44.6} &  {46.6} & 2.91GB & {4.53} & {4.97} & 264 &  {3.69hr} & {286} & 0.61 \\ 
  {3D Vehicle} & $\mathsf{R-A}$ & 15,435 & 99 & 0 & 1,528,065 & {1,004} & {1,340} & 92.6GB & {534} &  {551} & 36.3GB & {51.8}  & {46.5} & 3.3GB & {$>$24hr} & {5,933} & 1.22 \\
		3D RoomTemp & $\mathsf{S}$ & 9,261 & 36 & 0 & 333,396 & {1.51} & {78.5} & 24.7GB & - & - & - & {0.015} & {0.014} & 2.7 & {136$^{*}$} & {154$^{*}$} & 0.52 \\
		4D BAS & $\mathsf{S}$ & 1,225 & 4 & 0 & 4,900 & {0.89} & {1.33} & 48.02 & - & - & - & {0.004} & {0.007} & 0.04 & {6.37hr$^{*}$} & {3,038$^{*}$} & 0.07 \\
		5D RoomTemp & $\mathsf{S}$ & 7,776 & 36 & 0 & 279,936 & {1.2} & {167.6} & 17.4GB & - & - & - & {0.21} & {0.24} & 2.24 & {$97.88^{*}$} & {$111.5^{*}$} & 0.56 \\
		7D BAS & $\mathsf{S}$ & 107,163 & 0 & 0 & 107,163 & {1.47hr}& {2.03hr} & 91.9GB & - & - & - & {0.501} & {0.139} & 0.86 & {$>$24hr} & {$>$24hr} & 7.7 \\
		14D Case & $\mathsf{S}$ & 16,384 & 0 & 0 & 16,384 & {699} & {987} & 2.15GB & - & - & - & {0.041}& {0.201} & 0.13 & {623} & {67.7} & 2.1 \\
	\end{tabular}}\label{tab:time}
 \vspace{0.5cm}
\caption{\small
	Execution times for controller synthesis, comparing the solving of the linear program in~\eqref{eq:linear-program} using GNU Linear Programming Kit (GLPK) against the sorting method from~\cite[Lemma 7]{sen2006model}. The comparison  is conducted on both a CPU (Intel i9-12900) and a GPU (NVIDIA RTX A4000), with times reported in seconds. The symbol ``$*$'' denotes that a finite horizon of $10$ seconds was utilized for the case study synthesis.}
{\footnotesize \begin{tabular}{|c|c|c|c|c|c|c|c|c|c|c|}
		Case Study & Spec & \multicolumn{4}{|c|}{} & \multicolumn{1}{|c|}{GLPK Library} & \multicolumn{2}{|c|}{Sorted LP Method} \\
		\hline
		& &$\vert\hat{X}\vert$ &$\vert\hat{U}\vert$&$\vert\hat{W}\vert$&$\vert\hat{X}\times\hat{U}\times\hat{W}\vert$ & CPU i9 & CPU i9 & GPU RTX\\
		\hline
  {4D BAS$^*$} & {$\mathsf{S}$} & 1,225 & 4 & 0 & 4,900 & {23.2} & {0.38} & {0.53} \\
  {3D RoomTemp$^*$} & {$\mathsf{S}$} & 216 & 36 & 0 & 7,776 & {0.34} & {0.22} & {0.29} \\
		2D Robot & $\mathsf{R}$ & 441 & 121 & 0 & 53,361 & 13.07 & 2.26 & 5.2\\
      {5D RoomTemp$^*$} & {$\mathsf{S}$} & 7,776 & 9 & 0 & 69,984 & {160} & {22.2} & {76.2} \\
  {3D Vehicle} & {$\mathsf{R-A}$} & 7,938 & 30 & 0 & 238,140 & {$>$3.0hr} & {241} & {490} \\
		2D Robot & $\mathsf{R-A}$ & 1,681 & 441 & 0 & 741,321 & 1691 & 577 & 216 \\
\end{tabular}}\label{tab:GPU} \vspace{-3cm}
\end{sidewaystable}

\end{document}